\documentclass[runningheads]{llncs}
\usepackage{graphicx}
\usepackage{hyperref}
\usepackage[final]{changes} 
\usepackage{soul}
\usepackage{float}
\usepackage{cite}

\usepackage{multicol} 
\usepackage{multirow} 

\usepackage[lined, ruled,,linesnumbered,resetcount]{algorithm2e}

\usetikzlibrary{fit,calc}
\newcommand*{\tikzmk}[1]{\tikz[remember picture,overlay,] \node (#1) {};\ignorespaces}
\newcommand{\boxit}[1]{\tikz[remember picture,overlay]{\node[yshift=3pt,fill=#1,opacity=.25,fit={(A)($(B)+(.95\linewidth,.8\baselineskip)$)}] {};}\ignorespaces}
\colorlet{pink}{red!40}
\colorlet{light_blue}{cyan!60}

\usepackage{hyperref}

\newcommand{\remove}[1]{}

\newcommand{\denseSpace}{\renewcommand{\baselinestretch}{0.80}\normalsize}
\newcommand{\normalSpace}{\renewcommand{\baselinestretch}{1.0}\normalsize}

\newcommand{\lref}[1]{line~\NlSty{\ref{#1}}}

\newcommand{\llref}[2]{lines~\NlSty{\ref{#1}}--\NlSty{\ref{#2}}}

\SetKwBlock{Loop}{loop}{end loop}
\SetKwData{Int}{\textbf{int}}
\SetKwData{Bool}{\textbf{boolean}}
\SetKw{Array}{array}
\SetKw{Of}{of}
\SetKw{Const}{const}
\SetKw{Define}{define}
\SetKw{Init}{init}
\SetKw{Goto}{goto}
\SetKw{Await}{await}
\SetKw{True}{true}
\SetKw{False}{false}
\SetKw{Cont}{continue loop}
\SetKw{Local}{local}
\SetKw{Shared}{shared}

\usepackage{amsmath,amssymb,amsfonts}
\usepackage{textcomp}
\begin{document}
\title{Machine-Learning Based Objective Function Selection for Community Detection\thanks{This work was supported in part by the Cyber Security Research Center at Ben-Gurion University.}}
%
%
\author{Asa Bornstein\inst{1}\orcidID{0000-0001-9607-2233} \and
Amir Rubin\inst{2,3}\orcidID{0000-0001-5356-6786} \and
Danny Hendler\inst{3}\orcidID{0000-0001-7152-7828}}
\authorrunning{A. Bornstein et al.}
%

\institute{Ben-Gurion University of the Negev, Beersheba 8410501, Israel \and
Department of Computer Science \\
\email{asabor@post.bgu.ac.il}}


%
\maketitle              
%

\begin{abstract}
NECTAR, a Node-centric ovErlapping Community deTection AlgoRithm, presented in 2016 by Cohen et. al, chooses dynamically between two objective functions which function to optimize, based on the network on which it is invoked. This approach, as shown by Cohen et al., outperforms six state-of-the-art algorithms for overlapping community detection. In this work,  we present NECTAR-ML, an extension of the NECTAR algorithm that uses a machine-learning based model for automating the selection of the objective function, trained and evaluated on a dataset of 15,755 synthetic and 7 real-world networks. Our analysis shows that in approximately 90\% of the cases our model was able to successfully select the correct objective function. We conducted a competitive analysis of NECTAR and NECTAR-ML. NECTAR-ML was shown to significantly outperform NECTAR's ability to select the best objective function. We also conducted a competitive analysis of NECTAR-ML and two additional state-of-the-art multi-objective community detection algorithms. NECTAR-ML outperformed both algorithms in terms of average detection quality. Multiobjective EAs (MOEAs) are considered to be the most popular approach to solve MOP and the fact that NECTAR-ML significantly outperforms them demonstrates the effectiveness of ML-based objective function selection.

\end{abstract}

\section{Introduction}
\label{seq:Introdcution}

Social networks tend to exhibit community structure, that is, they are partitioned to sets of nodes called communities (a.k.a. clusters), each of which relatively densely-interconnected, with relatively few connections between different communities. Revealing the community structure underlying complex networks is a key problem with many applications that is the focus of intense research (see e.g. 
\cite{krogan2006global, flake2002self, Pizzuti2014Algorithms, King2004Protein,Ahn2010Link}). Numerous community detection algorithms were proposed (see e.g. \cite{le2013fast, Lancichinetti_2009, Creusefond2016Evaluation,Kun2015Detecting, Viamontes2011Compression,Blondel_2008, Newman2004Finding, Xie2012Towards, Gregory_2010, Adamcsek2006CFinder, Harenberg2014survey}). While research focus was initially on detecting disjoint communities, in recent years there is growing interest also in the detection of overlapping communities, where a node may belong to several communities.

Many community detection algorithms are guided by an objective function that provides a quality measure of the clusterings they examine in the course of their execution (see e.g. \cite {Gao2019conductance, 
Blondel_2008, 
tasgin2007community,
Pizzuti2008GA-Net, 
Gong2011Memetic}). Since exhaustive-search optimization of these functions is generally intractable (see e.g. \cite{Brandes2007Finding, 
Sima2006NP}), existing methods search for an approximation of the optimum and employ heuristic search strategies. A key example of such heuristic is Blondel et al.'s algorithm \cite {Blondel_2008}, also known as the Louvain method (LM). The method aims to maximize the \emph{modularity} objective function \cite{Viamontes2011Compression}  as it employs a greedy local search heuristic that iterates over all nodes, assigning each node to the community it fits most (as quantified by modularity). Modularity  assumes disjoint communities. Which objective functions should be used for overlapping community detection? According to what criteria should they be chosen? NECTAR, a Node-centric ovErlapping Community deTection AlgoRithm - presented in 2016 by Cohen et. al \cite {NECTAR2017Cohen}, generalized Blondel et al.'s method, so that it can be applied also to the overlapping case. NECTAR chooses dynamically between two objective functions which to optimize, based on the network on which it is invoked. The selection is made between $Q^E$, an extension of modularity for overlapping communities \cite{Chen2014ExtensionModularity}, and WOCC - Weighted Overlapping Community Clustering, an extension of WCC (Weighted Community Clustering) \cite {prat2012shaping} for overlapping community detection. NECTAR selects which objective function to use based on the rate of closed triangles (out of all possible triangles) in the graph. This approach, as shown by Cohen et al. \cite {NECTAR2017Cohen}, outperforms six state-of-the-art algorithms for overlapping community detection.

In this work, we show that by extending NECTAR, by taking into account multiple features of the graph for selecting the objective function to be used, we are able to optimize objective function selection by using a machine learning model. This model selects an objective function which maximizes the quality of the clusters computed by NECTAR. As possible commonly-used metrics can be used to quantify this quality, we generate an ML model per each such metric.

We analyzed 3,933 synthetic and 7 real-world networks, measuring the quality of our models, aiming to dynamically select, based on the properties of the graph at hand, which objective function should be used. Our analysis shows that in approximately 90\% of the cases our model was able to successfully select the correct objective function to maximize the desired metric.

\subsection{Our Contributions}
\label{seq:Our Contributions}

We present NECTAR-ML, a node-centric overlapping community detection algorithm which employs a machine learning model for the selection of the objective function. To train our models, we created a dataset of 15,755 synthetic networks, with various sizes and properties  \footnote{Using a parallel computing framework we developed, available at \href {https://github.com/asaborn/GenericMultiTasking}{https://github.com/asaborn/GenericMultiTasking}}. We conducted a competitive analysis of NECTAR and NECTAR-ML. NECTAR-ML was proven to be superior over NECTAR as it significantly outperformed NECTAR's ability to select the best objective function out of $Q^E$ and WOCC. In addition, we have conducted an extensive competitive analysis of NECTAR-ML and two additional state-of-the-art multi-objective community detection algorithms. NECTAR-ML outperformed both algorithms in terms of average detection quality.

The rest of this article is organized as follows. We survey key related work in Section \ref{seq:related}. We present the NECTAR-ML algorithm in Section \ref{seq:NECTAR-ML_Algorithm}. We report on our experimental evaluation in Section \ref{seq:ExperimentalEvaluation}. We conclude in Section \ref{sec:conc}.

\section{Related Work}
\label{seq:related}

Blondel et al.'s algorithm \cite {Blondel_2008}, a.k.a. the Louvain method, is a widely-used disjoint community detection algorithm. It is based on a simple  search heuristic that seeks to maximize \emph{modularity} \cite{Viamontes2011Compression} - a global objective function that estimates the quality of a graph partition to disjoint communites. Chen et al. extended the definition of modularity to the overlapping setting \cite {Chen2014ExtensionModularity} and their extended definition is denoted $Q^E$. Yang and Leskovec \cite {Yang2015ground-truth} conducted a comparative analysis of $13$ objective functions, for disjoint or overlapping communities, in order to determine which captures better the community structure of a network. They show that which function is best depends on the network at hand. They also observe that objective functions that are based on triadic closure - identification of nodes which close a triangle with other nodes in the graph, provides the best results when there is significant overlap between communities. Weighted Community Clustering (WCC) \cite {prat2012shaping}  is such an objective function, defined only for disjoint community structures.

NECTAR, a Node-centric ovErlapping Community deTection AlgoRithm, presented in 2016 by Cohen et. al \cite {NECTAR2017Cohen}, generalized Blondel et al.'s algorithm so it can be applied also to networks possessing overlapping community structure. As part of their work, Cohen et al. presented WOCC - Weighted Overlapping Community Clustering, a generalization of WCC to the overlapping case. In addition to WOCC, NECTAR also employed the $Q^E$ objective function. A unique feature of NECTAR is that it chooses dynamically whether to use WOCC or $Q^E$, depending on the structure of the graph in hand.
 NECTAR's decision as to which of the above objective functions to choose is based solely on the average number of closed triangles per node in the graph, denoted as $AverageTrianglesRate$. If the $AverageTrianglesRate$ value is below a certain threshold, $Q^E$ will be employed, otherwise WOCC will be used. This approach, as shown by Cohen et al., provided good separation between communities with high overlap (on which WOCC, in most cases, is superior) and low overlap (on which extended modularity, in most cases, is superior). An extensive experimental evaluation was conducted, comparing NECTAR and six other state-of-the-art overlapping community detection algorithms. The evaluation was done using both synthetic and real-world networks with ground-truth. The evaluation of the clusterings output by the algorithms was made using the commonly-used metrics ONMI \cite {Lancichinetti_2009}, Omega Index \cite {Linda1988Omega} and Average F1 score \cite {Yang2012Affiliation}. NECTAR outperformed all other algorithms in terms of average detection quality. It was ranked first (on average) for both synthetic and real-world networks, leading in 33 out of 96 of synthetic networks and was best for one real-world network and second-best for the other one.

Another approach for the problem of approximating the optimum of the objective function is by employing evolutionary algorithms(EAs) (see e.g. \cite{tasgin2007community, Pizzuti2008GA-Net, Gong2011Memetic,Girvan7821, Lancichinetti2009comparative, Clauset2008Hierarchical, Gong2014Swarm, Pizzuti2012Genetic, Chuan2012complex, GONG2012evolutionary, Liu2014Signed, Tian2020Fuzzy, Wen2017MaximalClique}). EAs are a class of optimization heuristics methods inspired by biological evolution. Candidate solutions to the optimization problem play the role of individuals in a population, and a fitness function is used to determine the quality of the solutions. A common approach is to consider the community detection problem as a single-objective optimization problem (see e.g. \cite{tasgin2007community, Pizzuti2008GA-Net, Gong2011Memetic} ). However, it is plausible to assume that a community should have dense intraconnections and sparse interconnections, implying that two objectives should be optimized simultaneously in community detection, i.e., maximizing internal links and minimizing external links \cite{Girvan7821, Lancichinetti2009comparative, Clauset2008Hierarchical}. Therefore, the community detection problem can also be modelled as a multiobjective optimization problem (MOP). 
	Multiobjective EAs (MOEAs) are considered to be the most popular approach to solve MOP and indeed, several MOEAs \cite{Gong2014Swarm, Pizzuti2012Genetic, Chuan2012complex, GONG2012evolutionary,Liu2014Signed} have been proposed for the purpose of community detection. A unique feature of such algorithms is that they find a set of optimal solutions instead of a single solution in one run, as each solution corresponds to a partition of the given network. Most of the existing works focus on developing MOEAs for detecting nonoverlapping communities, and there are only a few  works (see e.g. \cite{Tian2020Fuzzy, Wen2017MaximalClique}) that focused on detecting overlapping communities. The reason for that is the challenge of the individual representation for encoding and decoding of overlapping communities. This challenging representation has a significant effect on MOEAs' scalability to large-scale networks in the real-world, causing them to be impractical for graphs containing over 10,000 nodes. 

Among the few MOEAs which are able to detect overlapping communities, we focus on two leading algorithms. The two differ in several aspects and present different types of representation of the nodes, as well as different mechanisms of biological evolution during the execution of the algorithm. One is the Maximal Clique based Multi Objective Evolutionary Algorithm (MCMOEA), introduced by Wen et al. \cite{Wen2017MaximalClique} in 2017. Wen et al. presented the maximal-clique graph by using a set of maximal cliques as nodes and the links among maximal cliques as edges. Then, based on the maximal-clique graph, MCMOEA uses the list of cliques, to which a node belongs, to represent the nodes in a topological manner. Since two maximal cliques are allowed to share the same nodes of the original graph, overlap is an intrinsic property of the nodes of the maximal clique graph. MCMOEA was compared to five state-of-the-art methods for overlapping community detection which include two MOEA algorithms.  This comparison confirmed that MCMOEA is competitive and yields promising results. One of the main reasons that we have selected the MCMOEA algorithm as one of the algorithms with which we compare NECTAR-ML, is its ability to process networks of size up to 10,000 nodes. The second algorithm is the Evolutionary Multiobjective Optimization based Fuzzy Method (EMOFM), introduced by Tian et al.'s \cite{Tian2020Fuzzy} in 2019. In fuzzy clustering, a node belongs to a community with a certain probability. This makes fuzzy clustering an effective technique for overlapping community detection, as it allows a node to belong to more than one cluster \cite{Gregory2011Fuzzy}. The crucial step of fuzzy clustering based overlapping community detection is to find the optimal community center for each cluster. The proposed method adopts the idea of fuzzy c-medoids \cite{Krishnapuram2001fuzzy}, where each community center is a node in the network. These nodes are optimized by a well tailored MOEA, and the number of communities can also be automatically determined. Moreover, the MOEA is also used to optimize the fuzzy threshold of each node, which enables the proposed method to find diverse overlapping community structures. While in \cite{Tian2020Fuzzy} three approaches of EMOFM are presented, in this paper, we use its best approach in terms of performance and run time compared to the other two, denoted EMOFM-DK. DK stands for the use of Diffusion Kernel similarity for measuring the distance between nodes in the graph \cite {Kondor2002Diffusion}. EMOFM-DK outperformed six state-of-the-art approaches for overlapping community detection, including MCMOEA and other two MOEA algorithms.

Both the MCMOEA and the EMOFM-DK algorithms use two exact metrics for quantitatively comparing the quality of overlapping communities obtained by different approaches on the benchmark networks, i.e., the extended modularity $Q^E$ and the generalized normalized mutual information (gNMI) \cite {Lancichinetti_2009}. Similarly to ONMI, which is used as an evaluation criteria in our work, gNMI is also a version of Normalized Mutual Information (NMI), suitable for overlapping clustering evaluation. Both evaluation criteria belong to the family of Information Theory Based Metrics, described in \cite{Lutov2019Evaluation}. While $Q^E$ can be used without knowing the true community structure, gNMI is used only for networks whose ground truth is known.

Both Wen at el. and Tian et al. executed 30 independent runs for each network. For each run, the best $Q^E$ and gNMI values are recorded (out of all  $Q^E$ and gNMI values, calculated for each clustering output). 

According to Wen et al., although MCMOEA shows great superiority in terms of gNMI over other algorithms, recent studies \cite{Romano2014Standardized, Kondor2002Diffusion, Amelio2015Normalized, mcdaid2013normalized, Vinh2010Information} have shown that gNMI may suffer from the selection bias problem that tends to choose solutions with more communities. For this reason, Wen et al. have also used FNMI\cite{Amelio2015Normalized} as an additional evaluation criteria.

We emphasize that the main difference between NECTAR's approach to that of MOEA algorithms is the fact that NECTAR is striving to select the best objective function among multiple objective functions for the optimization problem, while the MOEA algorithms are simultaneously optimizing multiple objective functions. Another important difference between the two approaches is that NECTAR was designed to handle large-scale networks, while MOEA algorithms are capable of processing networks comprised of at most a few thousand nodes. 
\section{NECTAR-ML Algorithm}
\label{seq:NECTAR-ML_Algorithm}

The high-level pseudo-code of the NECTAR-ML algorithm is given by Algorithm \ref{alg:NECTAR-ML} alongside NECTAR's original code, given by Algorithm \ref{alg:NECTAR}. As described in Section \ref{seq:related}, NECTAR's decision as to which objective functions to choose, is based on the average number of closed triangles per node in the graph (\llref{ifHighTriRate}{useEmod} in Algorithm \ref{alg:NECTAR}, coloured in \textcolor{pink}{pink}). NECTAR-ML optimizes this selection using a machine learning model (\llref{NECTAR-ML:ExtractFeatures}{NECTAR-ML:UseObjFunc} in  Algorithm \ref{alg:NECTAR-ML}, coloured in \textcolor{blue}{blue}). As the rest of the code is the same in both algorithms, in section \ref{seq:DetailedDescription}, we use the text describing NECTAR in \cite {NECTAR2017Cohen} for describing this part of NECTAR-ML's pseudo-code.

\subsection{Detailed Description}
\label{seq:DetailedDescription}

The input to the \texttt{NECTAR-ML} procedure (see \lref{NECTAR-ML:start}) is a graph $G=<V,E>$ and an algorithm parameter $\beta \geq 1$ that is used to determined the number of communities to which a node should belong in a dynamic manner (as we  describe below).

\texttt{NECTAR-ML} proceeds in iterations (\llref{NECTAR-ML:iterLoopStart}{NECTAR-ML:iterLoopEnd}), which we call \emph{external iterations}. In each external iteration, the algorithm performs \emph{internal iterations}, in which it iterates over all nodes $v \in V$ (in some random order), attempting to determine the set of communities to which node $v$ belongs such that the objective function is maximized. NECTAR-ML extracts features from the graph $G$ and provides them as input to the model, which then predicts which of the objective functions, WOCC or $Q^E$, is best to use (\llref{NECTAR-ML:ExtractFeatures}{NECTAR-ML:UseObjFunc}).

Each internal iteration (comprising \llref{NECTAR-ML:nodesDo}{NECTAR-ML:nodesDoEnd}) proceeds as follows. First, \texttt{NECTAR-ML} computes the set $C_v$ of communities to which node $v$ currently belongs (\lref{NECTAR-ML:computeCv}). Then, $v$ is removed from all these communities (\lref{NECTAR-ML:removeFromCv}). Next, the set $S_v$ of $v$'s neighboring communities (that is, the communities of $\cal{C}$ that contain one or more neighbors of $v$) is computed in \lref{NECTAR-ML:Sv}. Then, the gain in the objective function value that would result from adding $v$ to each neighboring community (relative to the current set of communities $\cal{C}$) is computed in \lref{NECTAR-ML:computeDeltas}. Node $v$ is then added to the community maximizing the gain in objective function and to any community for which the gain is at least a fraction of $1/\beta$ of that maximum (\llref{NECTAR-ML:computeC'v}{NECTAR-ML:addToC'v}).\footnote{If no gain is positive, $v$ remains as a singleton.}
Thus, the number of communities to which a node belongs may change dynamically throughout the computation, as does the set of communities $\cal{C}$.

If the internal iteration did not change the set of communities to which $v$ belongs, then $v$ is a \emph{stable node} of the current external iteration and the number of stable nodes (which is initialized to $0$ in \lref{NECTAR-ML:initStable}) is incremented (\llref{NECTAR-ML:noChange}{NECTAR-ML:incrementStable}). After all nodes have been considered, the possibly-new set of communities is checked in order to prevent the emergence of different communities that are too similar to one another. This is accomplished by the \texttt{merge} procedure (whose code is not shown), called in \lref{NECTAR-ML:merge}. It receives as its single parameter a value $\alpha$ and merges any two communities whose relative overlap is $\alpha$ or more. More precisely, each pair of communities $C_1, C_2 \in \cal{C}$ is merged if $\vert C_1 \cap C_2 \vert / min\{|C_1|,|C_2|\} \geq \alpha$ holds. We use $\alpha=0.8$, as this is the value that gave the best results (\lref{Var:alpha}). If the number of communities was reduced by \texttt{merge}, the counter of stable nodes is reset to $0$ (\llref{NECTAR-ML:Ifmerged}{NECTAR-ML:resetS}).

The computation proceeds until either the last external iteration does not cause any changes (hence the number of stable nodes equals $\vert V \vert$) or until the maximum number of iterations is reached (\lref{NECTAR-ML:RepeatCondition}), whichever occurs first. We have set the maximum number of iterations to $20$ (\lref{Var:maxIter}) in order to strike a good balance between detection quality and runtime. In practice, the algorithm converges within a fewer number of iterations in the vast majority of cases.  For example, in \cite {NECTAR2017Cohen}, Cohen et. al conducted experiments on synthetic graphs with $5000$ nodes, resulting in NECTAR's convergence after at most $20$ iterations in $99.5$\% of the executions.

\denseSpace
\NoCaptionOfAlgo
\begin{algorithm}[h]
\DontPrintSemicolon
\caption{\textbf{Algorithm 1} NECTAR algorithm pseudo-code. \label{alg:NECTAR}}
{\fontsize{9}{9}\selectfont
\Const $\text{maxIter}$ $\gets$ 20 \tcc*{max iterations} \nllabel{Var:maxIter}
\Const $\alpha$ $\gets$ 0.8 \tcc*{merge threshold} \nllabel{Var:alpha}
\tikzmk{A}
\Const $trRate$ $\gets$ 5 \nllabel{Var:trRate} \tcc*{WOCC threshold}
\tikzmk{B}
\boxit{pink}
\BlankLine

\textbf{Procedure} \texttt{NECTAR}(\emph{G=$<$V,E$>$}, $\beta$)\{ \nllabel{NECTAR:start}\;
\tikzmk{A}
\uIf{$triangles(G)/\vert V \vert \geq  trRate$ \nllabel{ifHighTriRate}}
    {use WOCC \tcc*{use WOCC obj. function}}
\Else{use $Q^E$ \tcc*{use $Q^E$ obj. function} \nllabel{useEmod}}
\tikzmk{B}
\boxit{pink}
Initialize communities \nllabel{NECTAR:init}\\
$i \gets 0$ \tcc*{number of extern. iterations}

\Repeat{$(s = \vert V \vert) \lor (i = maxIter)$ \nllabel{NECTAR:RepeatCondition}}
    {\nllabel{NECTAR:iterLoopStart}
    $s \gets 0$ \tcc*{number of stable nodes} \nllabel{NECTAR:initStable}
    \ForAll{$v \in V$ \nllabel{NECTAR:nodesDo}}
        {
        $C_v$ $\gets$ communities to which $v$ belongs \nllabel{NECTAR:computeCv}\\
        Remove $v$ from all the communities of $C_v$ \nllabel{NECTAR:removeFromCv}\;
        $S_v \gets \{C \in {\cal{C}} \big{\vert} \exists u: u \in C \land (v,u) \in E \}$ \nllabel{NECTAR:Sv}\;
        $D \gets \{\Delta(v,C) \vert C \in S_v \}$ \nllabel{NECTAR:computeDeltas}\;
        $C'_v \gets \{C \in S_v \vert \Delta(v,C) \cdot \beta \geq max(D) \}$ \nllabel{NECTAR:computeC'v}\;
        Add $v$ to all the communities of $C'_v$ \nllabel{NECTAR:addToC'v}\;
        \uIf{$C'_v = C_v$ \nllabel{NECTAR:noChange}}
            {
            $s$++ \nllabel{NECTAR:incrementStable}
            }
        } \nllabel{NECTAR:nodesDoEnd}
            \texttt{merge}($\alpha$) \tcc*{merge communities} \nllabel{NECTAR:merge}
            \uIf{\emph{merge reduced number of communities} \nllabel{NECTAR:Ifmerged}}
                {
                $s \gets$0 \nllabel{NECTAR:resetS}
                }
        $i$++\;
    } \nllabel{NECTAR:iterLoopEnd}}

\end{algorithm}
\normalSpace

\denseSpace
\NoCaptionOfAlgo
\begin{algorithm}[h]
\DontPrintSemicolon
\caption{\textbf{Algorithm 2} NECTAR-ML algorithm pseudo-code. \label{alg:NECTAR-ML}}
{\fontsize{9}{9}\selectfont
\Const $\text{maxIter}$ $\gets$ 20 \tcc*{max iterations} \nllabel{Var:maxIter}
\Const $\alpha$ $\gets$ 0.8 \tcc*{merge threshold} \nllabel{Var:alpha}

\BlankLine

\textbf{Procedure} \texttt{NECTAR-ML}(\emph{G=$<$V,E$>$}, $\beta$)\{ \nllabel{NECTAR-ML:start}\;
\tikzmk{A}
\tcc{Extract features from Graph} 
$features$ $\gets$ \texttt{ExtractFeatures}(G)  \nllabel{NECTAR-ML:ExtractFeatures} \\
\tcc{Predict obj. function}
$objFunc$ $\gets$ \texttt{model.predict($features$)} \nllabel{NECTAR-ML:PredictObjectiveFunc}\\
use $objFunc$ as the objective function \nllabel{NECTAR-ML:UseObjFunc} \\
\tikzmk{B}
\boxit{light_blue}
Initialize communities \nllabel{NECTAR-ML:init}\\
$i \gets 0$ \tcc*{number of extern. iterations}

\Repeat{$(s = \vert V \vert) \lor (i = maxIter)$ \nllabel{NECTAR-ML:RepeatCondition}}
    {\nllabel{NECTAR-ML:iterLoopStart}
    $s \gets 0$ \tcc*{number of stable nodes} \nllabel{NECTAR-ML:initStable}
    \ForAll{$v \in V$ \nllabel{NECTAR-ML:nodesDo}}
        {
        $C_v$ $\gets$ communities to which $v$ belongs \nllabel{NECTAR-ML:computeCv}\\
        Remove $v$ from all the communities of $C_v$ \nllabel{NECTAR-ML:removeFromCv}\;
        $S_v \gets \{C \in {\cal{C}} \big{\vert} \exists u: u \in C \land (v,u) \in E \}$ \nllabel{NECTAR-ML:Sv}\;
        $D \gets \{\Delta(v,C) \vert C \in S_v \}$ \nllabel{NECTAR-ML:computeDeltas}\;
        $C'_v \gets \{C \in S_v \vert \Delta(v,C) \cdot \beta \geq max(D) \}$ \nllabel{NECTAR-ML:computeC'v}\;
        Add $v$ to all the communities of $C'_v$ \nllabel{NECTAR-ML:addToC'v}\;
        \uIf{$C'_v = C_v$ \nllabel{NECTAR-ML:noChange}}
            {
            $s$++ \nllabel{NECTAR-ML:incrementStable}
            }
        } \nllabel{NECTAR-ML:nodesDoEnd}
            \texttt{merge}($\alpha$) \tcc*{merge communities} \nllabel{NECTAR-ML:merge}
            \uIf{\emph{merge reduced number of communities} \nllabel{NECTAR-ML:Ifmerged}}
                {
                $s \gets$0 \nllabel{NECTAR-ML:resetS}
                }
        $i$++\;
    } \nllabel{NECTAR-ML:iterLoopEnd}}

\end{algorithm}
\normalSpace

\subsection{Learning a Model for Objective Function Selection}
\label{seq:TheModel}

For a given graph, we can either use WOCC or $Q^E$ as our objective function. Therefore, we need to construct a binary machine learning classifier  which, given a network, selects between the two. Towards this goal, we constructed a large labeled dataset of networks for the training and evaluation of our model. A detailed description of the data collection process is presented in section \ref{DataCollection}. As for model evaluation, since our key goal is for the model to select an objective function that will optimize a specific quality metric, we selected to use the Accuracy and Recall scores per class as the performance evaluation metrics of the model. We consider several supervised ML algorithms for generating the model, including decision trees, deep Learning and linear classifiers. We use the default threshold of 0.5 to select which objective function to use. 

As described in Section \ref{DataCollection}, the training dataset is unbalanced. Therefore, we use the Balanced Accuracy metric (see Equation \eqref{eq:ba}) which is used to estimate models using an average of recall obtained on each class.

\begin{equation}
\label{eq:ba}
BA = \frac{TPR + TNR}{2}.
\end{equation}

In order to improve the models' accuracy, we tune the hyperparameters according to the average of the Balanced Accuracy metric over 5-folds of the data, as described in section \ref{sec:ModelEvaluationandResults}.

\subsubsection{Data Generation}
\label{DataCollection}

Lancichinetti, Fortunato and Radicchi et al.  \cite{Lancichinetti2008Benchmark} introduced a set of benchmark graphs (henceforth the LFR benchmark) that provide heterogeneity in terms of node degree and community size distributions, as well as control of the degree of overlap between the ground truth communities. This benchmark is widely used in the field of overlapping community detection research (see e.g. \cite{Xie2012Towards, Gregory_2010, Gregory2011Fuzzy, NECTAR2017Cohen, Xie2013Comparative, lee2010detecting, Wen2017MaximalClique, Tian2020Fuzzy}). We used  LFR to generate a dataset of 15,755 networks, extending the parameters used by \cite{Xie2013Comparative} as follows: The total number of nodes, $n$, is enlarged to 10K-65K, the average node degree, $k$, is set to 10,20,40,60 or 80. The number of overlapping nodes, $O_n$, is set to 10\%, 25\%, 50\% for networks with an average node degree of 10,20 or 40 and, in addition, a value of 75\% for networks with an average node degree of 60 or 80. The number of communities an overlapping node belongs to, $O_m$, is set to $\{2,\dots ,10\}$ (networks with low average degree are parameterized only with the lower portion of this range). The mixing parameter for the topology, $mut$, is set to values from the range $\{0.1,\dots ,0.5\}$ (networks with low average degree are parameterized only with the lower portion of this range). The maximum node degree $maxK$ is set to 50 for the networks with low average degree and 100 or 120 for the networks with higher degrees. {The exponent for degrees distribution, $\tau_1$, and the exponent for community size distribution, $\tau_2$, were set to 2 and 1 respectively and they are constant throughout the experiments.} See table \ref{tab:synthticNetworks} in the Appendix for the full list of parameter values. For each combination of parameters we generated two distinct graph instances.

Our goal is to construct a classifier that uses structural features of the network to select which objective function to apply. The features we extracted for this purpose are:

\begin{enumerate}
  \item $GCC$ (Global clustering coefficient): \newline
  $\frac{3 \times Number Of Triangles}{Number Of Triplets}$, where a triplet consists of three nodes that are connected by either two (open triplet) or three (closed triplet) undirected edges. The expression $3 \times Number Of Triangles$ can also be referred as to the total number of closed triplets in the graph, as one triangle contains three closed triplets. 
  \item $ACC$ (Average clustering coefficient):  \newline
  $\frac{\sum_{u \in G}C_u}{Number Of Nodes}$, where $Cu$ is defined as \[
    Cu= 
\begin{cases}
    \frac{2 \times \vert E_v,_w \vert}{k_u \times (k_u - 1) },& \text{if } k_u > 1\\
    0,              & \text{otherwise}, 
\end{cases}.  
\] where $E_v,_w$ is the set of edges among node $u$'s neighbours and $k_u$ is the degree of $u$
  \item $RatioOfNodesInTriangle$ : \newline
  $\frac{Number Of Nodes In Triangles}{Number Of Nodes}.$
  \item $AverageNodeDegree$ :  
  $\frac{2 \times Number Of Edges}{Number Of Nodes}.$
  \item $AverageTrianglesRate$ :  
  $\frac{Number Of Triangles}{Number Of Nodes}.$
\end{enumerate}


As for the labels, we invoked NECTAR on each network, once using the WOCC objective function and once using the $Q^E$ objective function with 10 different values of $\beta \in  \{1.01,1.05,1.09,1.1,1.2,1.3,1.4,1.6,1.8,2.0\}$. Using the ground truth of the those networks, we evaluated NECTAR's outputs per network and $\beta$ value in terms of several commonly-used metrics  which served as the evaluation criteria in \cite {NECTAR2017Cohen} for the NECTAR algorithm.

The following provides the description of those metrics as described by Cohen et al in \cite {NECTAR2017Cohen}:

\begin{enumerate}

\item \emph{Overlapping Normalized Mutual Information} (ONMI) \cite{Lancichinetti_2009} is based on the notion of normalized mutual information and uses entropy to quantify the extent by which we may learn about one cover given the other and is defined as follows.
$$ONMI({\cal{C}}_1,{\cal{C}}_2) = 1- \frac{1}{2}(H({\cal{C}}_1|{\cal{C}}_2) + H({\cal{C}}_2|{\cal{C}}_1)),$$
where $H({\cal{C}}_1|{\cal{C}}_2)$ is the conditional entropy of cover ${\cal{C}}_1$ w.r.t. cover ${\cal{C}}_2$. As mentioned in \cite{mcdaid2013normalized}, in cases where one cover contains many more communities than the other, ONMI is not a good representation of a cover's quality. We will address this matter when quantifying the quality of a cover using the ground-truth for real-world networks.

\item \emph{Omega-index} \cite {Linda1988Omega} is based on the fraction of pairs that occur together in the same number of communities in both covers, with respect to the expected value of this fraction in the null model. Unlike ONMI, this measure refers to the nodes and the relationships between them, giving us a different view on a cover's quality w.r.t. ground-truth. We use the following version of Omega-index, used in \cite{Xie2013Comparative}.\\
$\omega({\cal{C}}_1,{\cal{C}}_2) = \frac{\omega_ u({\cal{C}}_1,{\cal{C}}_2)- \omega_ e({\cal{C}}_1,{\cal{C}}_2)}{1-\omega_ e({\cal{C}}_1,{\cal{C}}_2)}.$\\
$\omega_ u({\cal{C}}_1,{\cal{C}}_2) = \frac{1}{\binom{n}{2}} \sum\limits_{j=0}^{min(\vert {\cal{C}}_1 \vert,\vert {\cal{C}}_2 \vert)} |t_j({\cal{C}}_1) \cap t_j({\cal{C}}_2)|.$\\
$\omega_ e({\cal{C}}_1,{\cal{C}}_2) = \frac{1}{\binom{n}{2}^2} \sum\limits_{j=0}^{min(\vert {\cal{C}}_1 \vert,\vert {\cal{C}}_2 \vert)} |t_j({\cal{C}}_1)|\cdot |t_j({\cal{C}}_2)|.$\\
$t_j({\cal{C}}) = \{(x,y): |\{C \in {\cal{C}} :x,y \in C\}| = j\}.$\\

\item \emph{Average F1 score} ($\bar{F_1}({\cal{C}}_1,{\cal{C}}_2)$), as presented in \cite{Yang2012Affiliation}: For each community in the ground-truth and in the evaluated cover, we find the community in the other cover with the highest F1 score (in terms of node community-membership), where $F1(C_1,C_2)$ is the harmonic mean of precision and recall between node-sets $C_1$, $C_2$. We then compute the average score for ground-truth communities and the average score for the evaluated cover and compute their average.
$precision(C_1,C_2)=\frac{|C_1 \cap C_2 |}{|C_1|}.$\\
$recall(C_1,C_2)=\frac{|C_1 \cap C_2 |}{|C_2|}.$\\
$H(a,b) = \frac{2 \cdot a \cdot b}{a+b}.$\\
$F_1(C_1,C_2) = H(precision(C_1,C_2),recall(C_1,C_2)).$\\
$F_1(C_1,{\cal{C}}) = max\{F_1(C_1,C_i) : C_i \in {\cal{C}}\}.$\\
And $\bar{F_1}({\cal{C}}_1,{\cal{C}}_2)$ is set to be:\\
$\frac{1}{2|{\cal{C}}_1|} \sum\limits_{C_i \in {\cal{C}}_1} F_1(C_i,{\cal{C}}_2) + \frac{1}{2|{\cal{C}}_2|} \sum\limits_{C_i \in {\cal{C}}_2} F_1(C_i,{\cal{C}}_1).$\\

\end{enumerate}

We then label the networks as follows. For each metric type and network, the objective function which provides the best metric score, over all $\beta$ values, is selected as the label of that network for that metric. In addition, a network is also labeled, in the same manner, according to  the average of the three metrics. Therefore, four different labels are generated (one per metric and a 4\textsuperscript{th} for their average scores).

It is possible that for some metric, the gap in scores between the result of NECTAR, when invoked using one objective function to the other, is very small. As we use these labels to train our model, we would like the model to reflect the fact that, in these cases, it is not that important which objective function is selected. To achieve this goal, we weighted each of the networks according to the gap between the two results, according to Equation \eqref{eq:weight}:
\begin{equation}
\begin{aligned}
\label{eq:weight}
Weight(G,m) & = \\ 
& \frac{\vert WOCCScore_m(G) - ModScore_m(G) \vert}{\max(WOCCScore_m(G), ModScore_m(G))}.    
\end{aligned}
\end{equation}

Where G denotes the network and m denotes the metric type. $WOCCScore_m(G)$ is the best metric score over all $\beta$ values for the network, using the WOCC objective function. $ModScore_m(G)$ is the equivalent term for the $Q^E$ objective function case.


Table \ref{tab:LabelsDistribution} presents the labels distribution, with and without consideration of the networks weight, for a total of 15,755 synthetic networks. The data shown in the table indicates that all metric types present a significant class imbalance (Approximately  80\% labeled as $Q^E$ and 20\% as WOCC). 

\begin{table}[!b]
\centering
\caption[Labels distribution for synthetic networks]{Objective Function labels distribution for synthetic networks of size 10K-65K with and without consideration of the networks weight.}
\label{tab:LabelsDistribution}
\begin{tabular}{| l | c | c | c | c |}
\hline
& \multicolumn{2}{c |}{Unweighted} & \multicolumn{2}{c |}{Weighted}\\
\cline{2-5}
Metric         &$Q^E$  &WOCC    &$Q^E$ &WOCC \\
\hline\hline

ONMI         	   &81\% 	&19\%   &74\%  &26\%   \\ \hline
Average F1         &81\% 	&19\%   &76\%  &24\%   \\ \hline
Omega-Index        &82\% 	&18\%   &78\%  &22\%   \\ \hline
Metrics Average    &80\% 	&20\%   &77\%   &23\%    \\ \hline

\end{tabular}
\end{table}

\subsubsection{Training and Evaluation}
\label{sec:ModelEvaluationandResults}

\noindent \textbf{Synthetic Networks:} We generated NECTAR-ML models and evaluated each of them according to the following metrics:  ONMI \cite {Lancichinetti_2009}, Omega Index \cite {Linda1988Omega}, Average F1 score \cite {Yang2012Affiliation} and their average value. All comprised datasets, one per metric, were partitioned into a training set and a test set in the same manner: Networks containing 10K-50K vertices, which constitute 75\% of the dataset (11,822/15,755), were used as the training set while networks containing 55K-65K vertices which constitute 25\% of the dataset (3,933/15,755), were used as the test set. The purpose of this setting is to validate that the models are robust, in terms of their ability to scale to the size of the network, as large scale networks share similar attributes, regardless of their size. This will imply that the models possess the ability to infer, with high accuracy, the suitable objective function for networks that are larger than those used for their training. As the data is imbalanced, we used oversampling in the training process.

\vspace{-3pt}

\begin{table}[!b]
\centering
\caption[Tested  values for Hyperparameters selection]{Tested values per supervised ML algorithm for optimized hyper-parameters selection.  }
\label{tab:HyperParameters}
\resizebox{\columnwidth}{!}{%
\begin{tabular}{| c | l | l | l | }
\hline
 Model  &Hyperparameter Type      &Tested values\\ 
       
\hline\hline
 $GBDT$ &Number of Estimators &100, 200, 250, 300, 350, 400  \\
  &Max. Depth  		   &3, 4, 5, 8, 10, 15, 20, 25, 30, 35, 40  \\
  &Min. Samples Split	   &2, 3, 4, 5, 10  \\
  &Min. Samples Leaf     &2, 4, 3, 5, 10  \\
  &Learning Rate &0.01, 0.025, 0.05, 0.075, 0.1, 0.15, 0.2  \\
    
\hline

$Extra Trees$ &Number of Estimators &100, 200, 250, 300, 350, 400  \\
  &Max. Depth  		   &3, 4, 5, 8, 10, 15, 20, 25, 30, 35, 40  \\
  &Min. Samples Split	   &2, 3, 4, 5, 10  \\z
  &Min. Samples Leaf     &2, 3, 5, 10  \\

 \hline
 
 $RanndomForest$ &Number of Estimators &100, 200, 250, 300, 350, 400  \\
  &Max. Depth  		   &3, 4, 5, 8, 10, 15, 20, 25, 30, 35, 40  \\
  &Min. Samples Split	   &2, 4, 3, 5, 10  \\
  &Min. Samples Leaf     &2, 3, 5, 10  \\
 \hline
 
 $LinearSVC$ &C &0.0001, 100, 0.01, 0.1, 10, 25, 50, 100  \\
  &Tolerance  		   &1e-10, 1e-9, 1e-8, 1e-7, 1e-6, 1e-4, 0.001, 0.01, 0.2, 1, 10, 20, 100  \\
\hline
 
 $Deep Learning$ &Number of Layers &2, 3, 4  \\
  &Number of nodes in first layer  		   &16, 32, 50, 64, 80, 100  \\
  &Epochs  		   &100, 250, 500, 1000, 2000  \\
  &Batch size  		   &32, 64, 128, 256, 512, 1024  \\
  &Optimizer  		   &'Adam', 'SGD' \\
 
\hline

\end{tabular}
}
\end{table}%

Table \ref {tab:HyperParameters} presents the various supervised ML algorithms and hyperparameters we considered for all models. To select the best hyper-params, we use 5-fold cross validation, keeping the ratio between classes equal in each fold. Finally, each trained classifier $C_m$ was applied to the corresponding test set fold. Table \ref{tab:SyntheticResults} summarizes the performance of the models in the setting described above. The 5\textsuperscript{th} column presents the averaged balanced accuracy results for the 5-folds cross-validation runs. The maximum standard deviation for a fold is less than 0.01. This indicates that the models are not overfitted. The remaining columns present the balanced accuracy and recall per class over the test-set. The 6\textsuperscript{th} column present the balanced accuracy result (BA) and the last two columns present the recall results for each of the objective functions. The balanced accuracy average achieved over the validation and test sets for all metrics is close to 90\%. In addition, the recall values are high, indicating that the models show no significant preference for neither of the classes, in spite of the class imbalance problem.


\begin{table}[!t]
\centering
\caption{Performance of the selected ML algorithms in terms of balanced accuracy on the synthetic networks dataset.
}
\label{tab:SyntheticResults}
\resizebox{\columnwidth}{!}{%
\begin{tabular}{| l | l | l | l | l | l | l | l |}
\hline
 Metric &Best  &Hyperparameter       &Opt. &BA                  &BA   &Recall - $Q^E$ & Recall - WOCC  \\ 
        &Algorithm   &Type 						   &value     &5-Folds       &Test Set     &Test Set          &Test Set\\

\hline\hline
  & &Number of Estimators &400  & & & & \\ 
  & &Max. Depth  		   &4  & & & & \\ 
 ONMI &$GBDT$ &Min. Samples Split	   &3   & 0.880 & 0.881 & 0.806 & 0.956 \\
  & &Min. Samples Leaf     &2 & & & & \\ 
  & &Learning Rate &0.1  & & & & \\
    
\hline

 & &Number of Estimators &400 & & & & \\
  & &Max. Depth  		   &4 & & & & \\
 Average F1 &$GBDT$ &Min. Samples Split	   &3  & 0.884  & 0.883 & 0.838 & 0.928\\
  & &Min. Samples Leaf     &2 & & & & \\ 
  & &Learning Rate &0.1 & & & & \\
 
\hline

  & &Number of Estimators &300 & & & & \\
 Omega-Index  &$Extra Trees$  &Max. Depth  		   &25 & 0.905  & 0.907 & 0.842  & 0.971\\
   & &Min. Samples Split	   &2  & & & & \\
   & &Min. Samples Leaf     &2 & & & & \\
 
\hline

   & &Number of Estimators &300 & & & & \\ 
 Metrics-Average  &$Extra Trees$ &Max. Depth  		   &40 &0.901 & 0.899 & 0.844 & 0.954\\
   & &Min. Samples Split	   &2 & & & & \\
   & &Min. Samples Leaf     &2 & & & & \\ 

\hline

\textbf{Overall Average} & & &&\textbf{0.893}  & \textbf{0.893} & \textbf{0.833} & \textbf{0.952} \\

\hline


\end{tabular}
}
\end{table}%

\noindent \textbf{Real World Networks:} To validate the quality of our model, we evaluated it on real world networks from the Stanford Large Network Dataset Collection \cite{snapnets}. Five undirected, unweighted networks from three different domains were considered: 
\begin{enumerate}
 \item \textbf{Co-product purchasing network (Amazon)}: Nodes represent products and edges are between commonly co-purchased products. Products from the same category are viewed as a ground-truth community.\item \textbf{Co-publishing network (DBLP)}: Nodes represent authors. Adjacent nodes represent authors with at least one shared publication. Ground-truth communities are defined as sets of authors who published in the same journal or conference.
\item \textbf{Social networks (LiveJournal, Friendster, Youtube)}: Nodes represent users and edges represent friendship between two users. Ground-truth communities are defined by node membership in user-created groups.
\end{enumerate}

In \cite{Yang2015ground-truth}, Yang and Leskovec rate the quality of the ground-truth communities using six scoring functions, such as modularity, conductance, and cut ratio. They rank ground-truth communities based on the average of their ranks over the six scores and maintain the $5,000$ top ground-truth communities per each network. These are the ground-truth communities provided as part of the datasets of \cite{snapnets}. 

Similarly to the analysis done in \cite{Harenberg2014survey, Creusefond2016Evaluation, Kun2015Detecting}, we removed all the nodes (and incident edges) that did not belong to any of the top 5000 communities. The resulting graphs and ground-truth communities were used to evaluate the models (one per metric). In addition, we used the Amazon and DBLP full networks for evaluation, as they were used for the real world competitive analysis in NECTAR's paper \cite {NECTAR2017Cohen}. A summary of the properties of these graphs is presented in Table \ref{tab:rwNetwroks}.

\begin{table}[b]
\centering
\caption{Properties of real-world networks used.} 
\label{tab:rwNetwroks}
\begin{tabular}{| l | c | c | c | c |}
\hline
\textbf{Sub-Graphs}  &  & & \\ 
\hline\hline
Network      &Number of vertices      &Number of  of edges  &Number of  of Clusters\\
\hline
Amazon          &16,716	 	  &48,739    &1,517  \\
DBLP     		&93,432		  &335,520   &4,961  \\
Youtube     	&39,841	      &224,235   &4,771  \\
LiveJournal     &84,438	      &1,521,988 &4,703  \\
Friendster  	&220,015		  &4,031,793 &4,914   \\

\hline
\textbf{Full Graphs} &  & & \\ 
\hline\hline
DBLP   &317,080  &1,049,866 &13,477 \\
Amazon    &334,863  &925,872 &75,149 \\
\hline
\end{tabular}
\end{table}%

As mentioned in \cite {mcdaid2013normalized}, the evaluation criteria we use return meaningful results when applied to a pair of covers (a cover is a set of communities generated for the network) of more-or-less the same size. As carried out in \cite {NECTAR2017Cohen}, we ensure this as follows. Let $\cal{G}$ be the set of ground-truth communities and let $\cal{C}$ be the cover computed by the algorithm. For each ground-truth community, we choose a community in $\cal{C}$ which most closely resembles it. In a more formal manner, for each ground-truth community $G \in \cal{G}$, we choose a single community $C=argmax\{F1(G,C') : C' \in \cal{C}\}$. This generates a subset ${\cal{D}} \subset \cal{C}$ of size not greater than $\vert \cal{G}\vert$. (There could be less communities in $\cal{D}$ than in $\cal{G}$ as duplicates are removed.) Then, we quantify the quality of $\cal{D}$ by computing the ONMI, Omega-index and average F1 score criteria.

Similarly to the synthetic networks, the labels of the real-world networks were assigned by applying NECTAR to each real-world network, applying both the WOCC and $Q^E$ objective functions with the 10 different values of $\beta$. Then, each network is labeled for the best objective function per metric. Table \ref{tab:rwWeightsAndHitRage} presents the labels for each metric and real-world network, including the network weight according to Equation \eqref{eq:weight}. It can be seen that the extended Modularity objective function is the preferable choice in 85\% of the cases (24/28) as the models for the ONMI and Average F1 metrics contain WOCC labels for the LiveJournal and Friendster networks while the other metrics models contain only $Q^E$ labels.

Table \ref{tab:rwWeightsAndHitRage} reveals the motivation behind splitting the analysis per metric, as some users might have different needs in evaluating the clustering results. For instance, from table \ref{tab:rwWeightsAndHitRage}, a user who chooses to use the Average-Metrics model with NECTAR-ML for the Friendster or LiveJournal social networks, will indeed receive better results for the average metrics result, but if the user is only interested in the ONMI or Average F1 metrics scores, it would be best for her to use the ONMI or Average F1 models for that purpose, as those models will provide the highest scores for the required metric.
 
\begin{table}[!t]
\centering
\caption{Ground truth labels and weights for real world networks. For each real world network, the best objective function and network weight are presented.
}
\label{tab:rwWeightsAndHitRage}
\begin{tabular}{| l | c | c | c | c | c | c | c | c |}
\hline
\textbf{Metric} & \multicolumn{2}{c|}{ONMI} & \multicolumn{2}{c|}{Average F1} &\multicolumn{2}{c|}{Omega-Index} &\multicolumn{2}{c|}{Metrics Average}\\
\hline
\textbf{Sub-Graphs} &Best  &Weight    &Best &Weight &Best &Weight  &Best &Weight   \\ 
\hline
Amazon          &$Q^E$	&0.074   &$Q^E$  &0.050  &$Q^E$ &0.273  &$Q^E$ &0.100 \\
Youtube         &$Q^E$	&0.235   &$Q^E$  &0.171  &$Q^E$ &0.401 &$Q^E$ &0.192 \\
DBLP            &$Q^E$	&0.012   &$Q^E$  &0.008  &$Q^E$ &0.226  &$Q^E$ &0.045  \\
LiveJournal     &WOCC	&0.038   &WOCC  &0.020  &$Q^E$ &0.619  &$Q^E$ &0.032  \\
Friendster      &WOCC	&0.003   &WOCC  &0.0002  &$Q^E$ &0.204  &$Q^E$ &0.056 \\

\hline
\textbf{Full Graphs} &      &  &  & &      &  &  & \\ 
\hline
DBLP           &$Q^E$	&0.071   &$Q^E$  &0.014  &$Q^E$ &0.695  &$Q^E$ &0.027  \\
Amazon         &$Q^E$	&0.127   &$Q^E$  &0.075  &$Q^E$ &0.259  &$Q^E$ &0.128  \\
\hline


\end{tabular}
\end{table}%
 
In Cohen et. al's NECTAR \cite {NECTAR2017Cohen}, the WOCC algorithm was presented as the superior selection by far for the full DBLP network, while in our experiments the results have shown that $Q^E$ is the superior objective choice. The reason for this difference is the fact that due to optimizations that we have introduced to NECTAR's code, higher values of $\beta$ could now be run with the $Q^E$ algorithm. While in Cohen et al's NECTAR\cite {NECTAR2017Cohen}, the maximum $\beta$ value which was used for $Q^E$ runs in NECTAR's competitive analysis was set to 1.4, in our experiments we used higher values as high as 2.0, which yielded a different set of results. For instance, the highest metric scores for the full DBLP network, for all metric types  (including the average of the three) except the Omega-Index metric, were produced using the $Q^E$ objective function and a $\beta$ value of 2.0. Scores produced with smaller $\beta$ values and the $Q^E$ objective function were not near the top ranked scores.

\begin{figure}[t]
\includegraphics[width = 0.6 \linewidth]{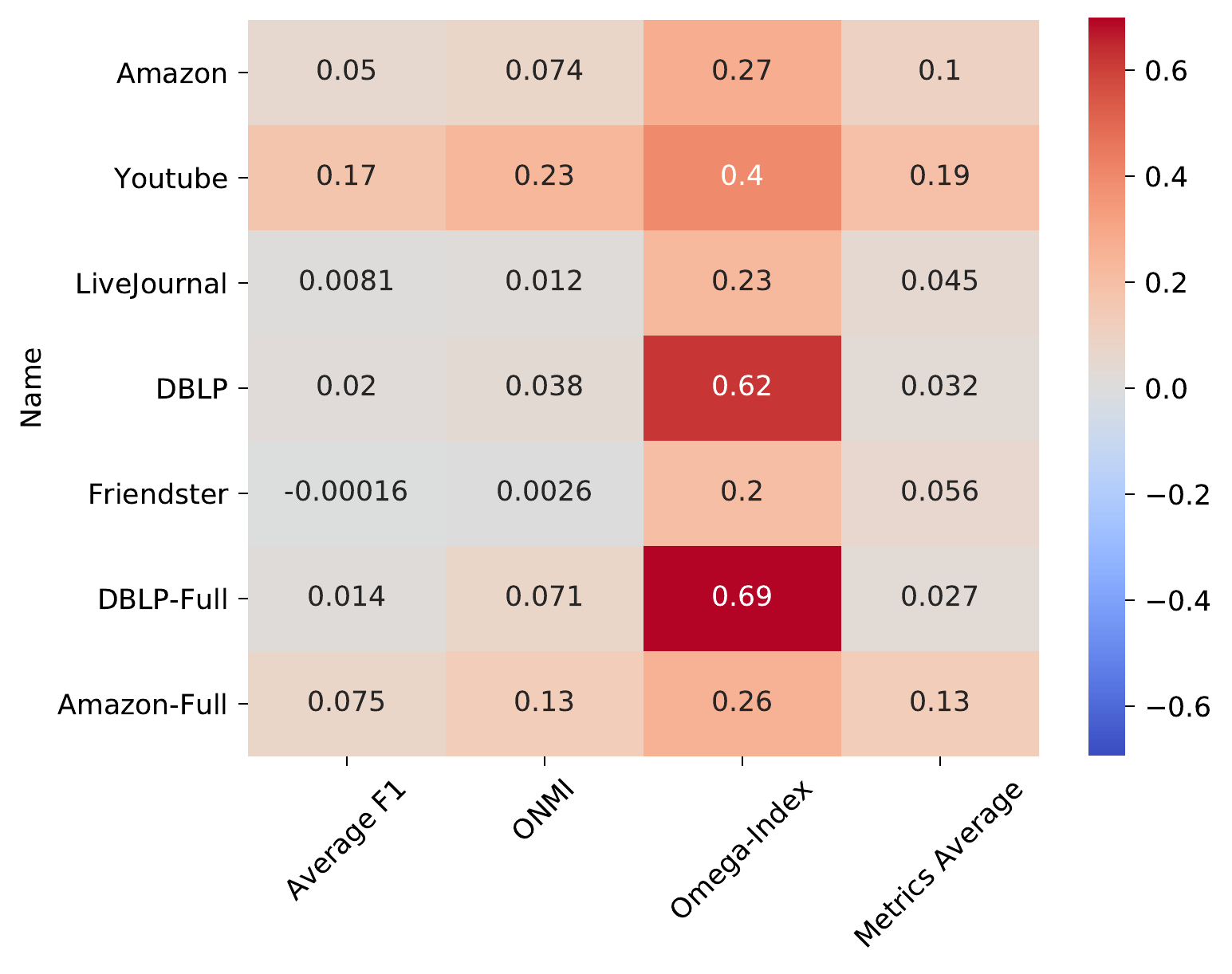}
\caption[A heat-map of the weighted hit/miss results on real world networks using the classifiers per metric]{A heat-map of the weighted hit/miss results on real world networks using the classifiers per metric. 
}
\label{fig:real_world_eval_heatamp} 
\end{figure}

In order to evaluate the models' performance on the real world networks, each classifier $C_m$ was trained using the selected hyperparameters on the whole synthetic networks dataset (10K-65K) and then applied on the real-world networks. Fig. \ref{fig:real_world_eval_heatamp} presents the weighted hit/miss results of the predictions of all classifiers. The values in the heatmap represent networks weight and signify the level of importance in selecting the objective function. A hit is represented by a positive value of the network weight and therefore coloured in red. A miss is represented by a negative value of the network weight and therefore coloured in blue. Weights close to zero are coloured in grey. By examining the weights values, we can see that the classifiers presented a high level of accuracy on both low and high weight values. It can also be seen that there is a single miss, by the Average F1 classifier, whose significance is very low (the weighted value is practically zero). Table \ref{tab:rwBAresults} summarizes the averaged results of the experiments conducted on the real-networks. The network weights are calculated using Equation \eqref{eq:weight} and presented in Table \ref{tab:rwWeightsAndHitRage}. The 2\textsuperscript{nd} column presents the averaged balanced accuracy results per metric while the remaining two columns present the averaged recall results for each of the objective functions. Recall values for WOCC are missing for the Omega-Index and Metrics-Average since no WOCC ground truth labels exist for those metrics.. The balanced accuracy weighted average of all metrics reaches nearly 100\%.

\begin{table}[!b]
\centering
\caption{Averaged test set results for real world networks with respect to the networks weights.
}
\label{tab:rwBAresults}
\resizebox{0.6 \columnwidth}{!}{%
\begin{tabular}{| l | c | c | c |}
\hline
Metric                   &Balanced accuracy        & Recall & Recall  \\ 
                   &        & $Q^E$  &WOCC  \\
\hline\hline

ONMI           & 1.000 & 1.000  & 1.000  \\
Average F1     & 0.990 & 1.000 & 0.980  \\
Omega-Index    & 1.000 & 1.000 & ---  \\
Metrics Average  & 1.000 & 1.000 & ---  \\
\hline

\textbf {Weighted Average}  & \textbf{0.999} & \textbf{1.000} &\textbf{0.999} \\
\hline


\end{tabular}}
\end{table}

\subsection{Feature Analysis}

The purpose of this section is to perform feature analysis for the features presented in section  \ref {DataCollection} with respect to the labels given for each network in the dataset of the 10K-65K networks sizes. We aim to explore the level of contribution each feature or a set of features has on the classification of the networks.

\begin{figure}[!t]
\includegraphics[width = 0.7 \linewidth]{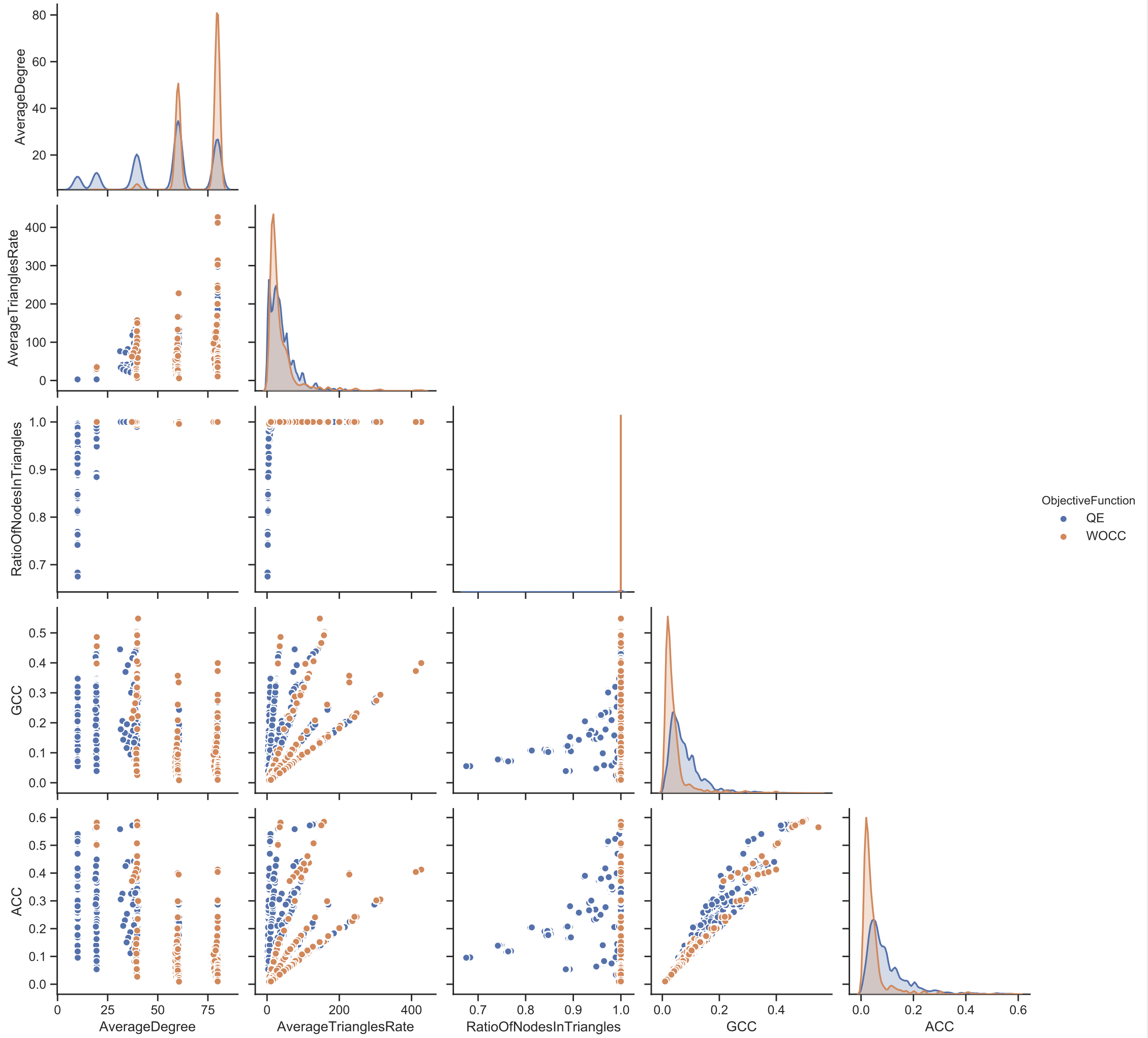}
\caption[Information Gain illustration for the average metrics model]{Information Gain illustration for the average metrics model.
}
\label{fig:IG_Average} 
\end{figure}

Fig. \ref {fig:IG_Average} presents the information gain analysis for the average metrics model. The graphs on the diagonal present a histogram of each class according to the values of a specific feature. The graphs under the diagonal presents a 2-D display of the distribution of the classes as a function of any pair of features. Information gain figures related to the ONMI, Omega-index and average F1 metrics carry the same trend and are presented in the Appendix (Figs. \ref{fig:IG_ONMI}-\ref{fig:IG_AverageF1}). In \cite {NECTAR2017Cohen}, NECTAR outperformed all the community detection algorithms, with which it is compared, using a simple selection method, which, as we can infer from Fig. \ref {fig:IG_Average}, can be improved significantly since there is no clear separation between $Q^E$ or WOCC based on $AverageTrianglesRate$.

\section{Experimental Evaluation}
\label{seq:ExperimentalEvaluation}

This section includes a comparison of NECTAR-ML to NECTAR and MOEA algorithms on synthetic and real world networks.

\subsection{Competitive Analysis to NECTAR}

In this section we present a competitive analysis between NECTAR-ML and NECTAR. We analyse the performance of the two methods on the test sets of both synthetic and real world networks, presented in section \ref {sec:ModelEvaluationandResults}. We compare the level of accuracy each algorithm provides, in selecting the best objective function for each of the test set networks. The specific model, per metric type, was applied on the test set networks and its objective function predictions are compared to NECTAR's dynamic selections according to the $AverageTriangleRate$ threshold value.

\begin{figure}[!t]
\centering
\includegraphics[width = \linewidth]{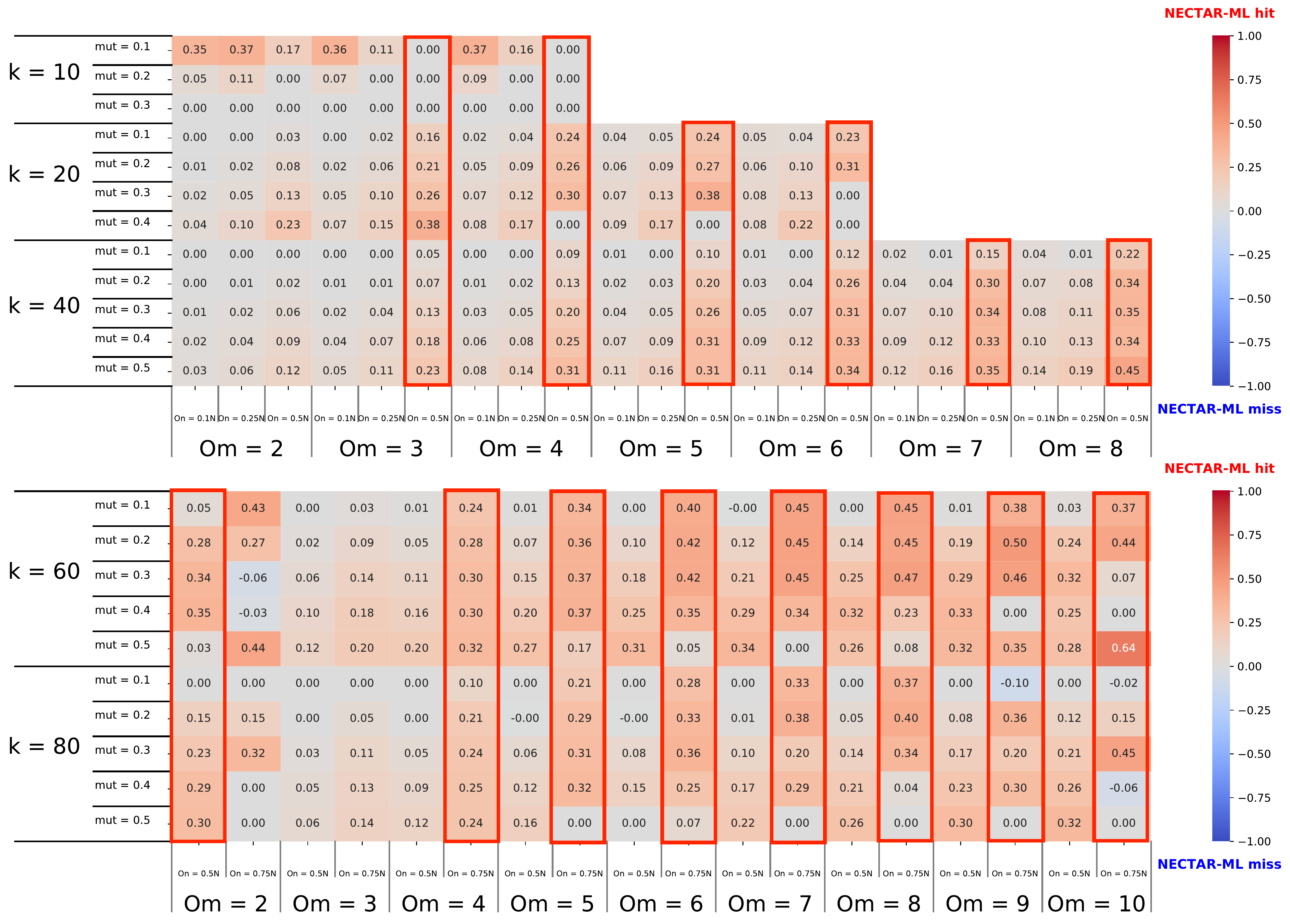}
\caption[NECTAR vs NECTAR-ML competitive analysis on synthetic networks using the average metrics model]{NECTAR vs NECTAR-ML competitive analysis on synthetic networks using the average metrics model. The comparison was made using the 55K-65K test set networks.
} 
 \label{fig:compAnalysis_nectar_nectar_prime} 

\end{figure}

\begin{figure*}
\centering
\includegraphics[width = \linewidth]{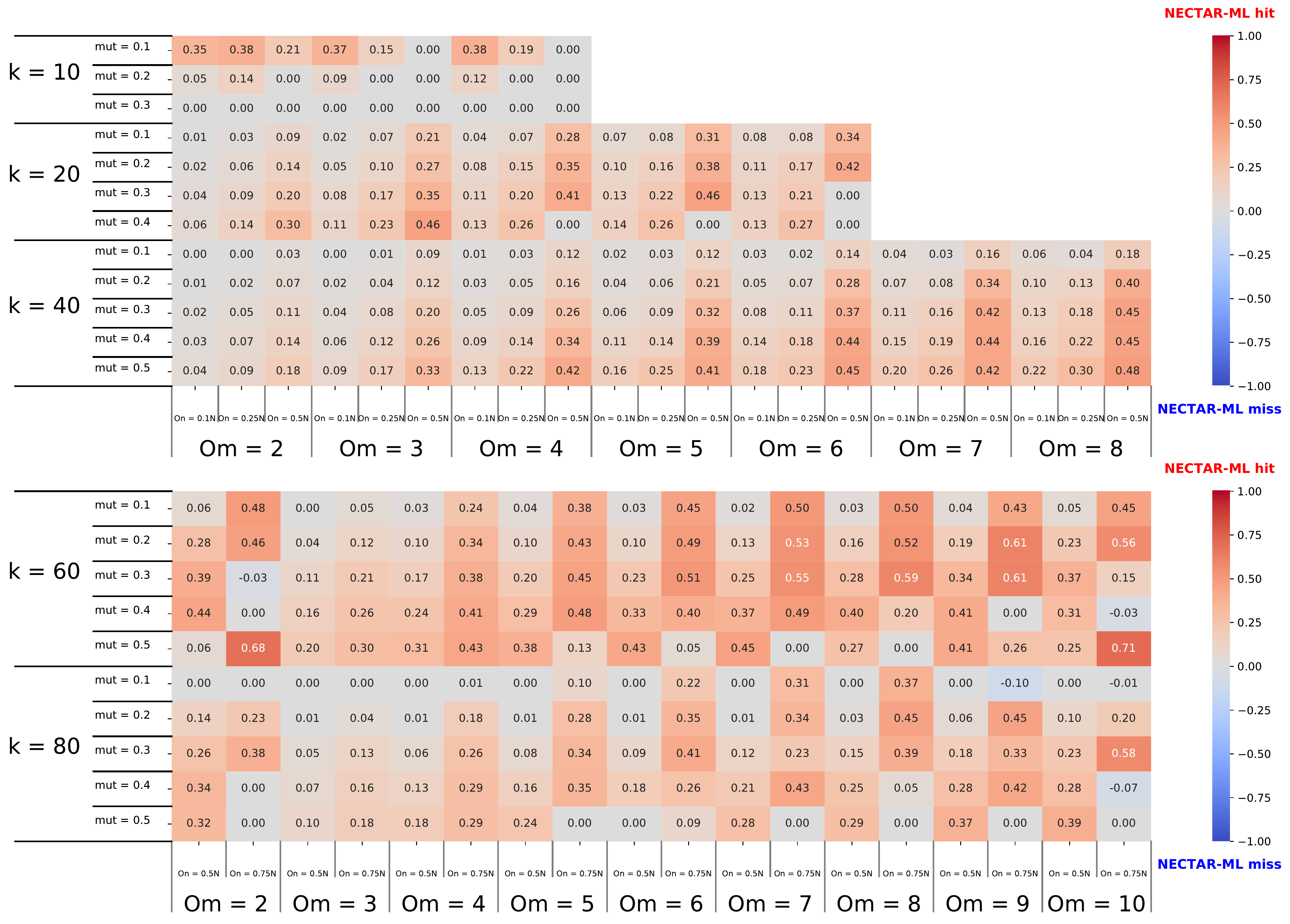}
\caption{NECTAR vs NECTAR-ML competitive analysis on synthetic networks using the Omega-Index metrics model.} 
 \label{fig:compAnalysis_nectar_nectar_ml_OmegaIndex} 
\end{figure*}

\begin{figure*}
\centering
\includegraphics[width = \linewidth]{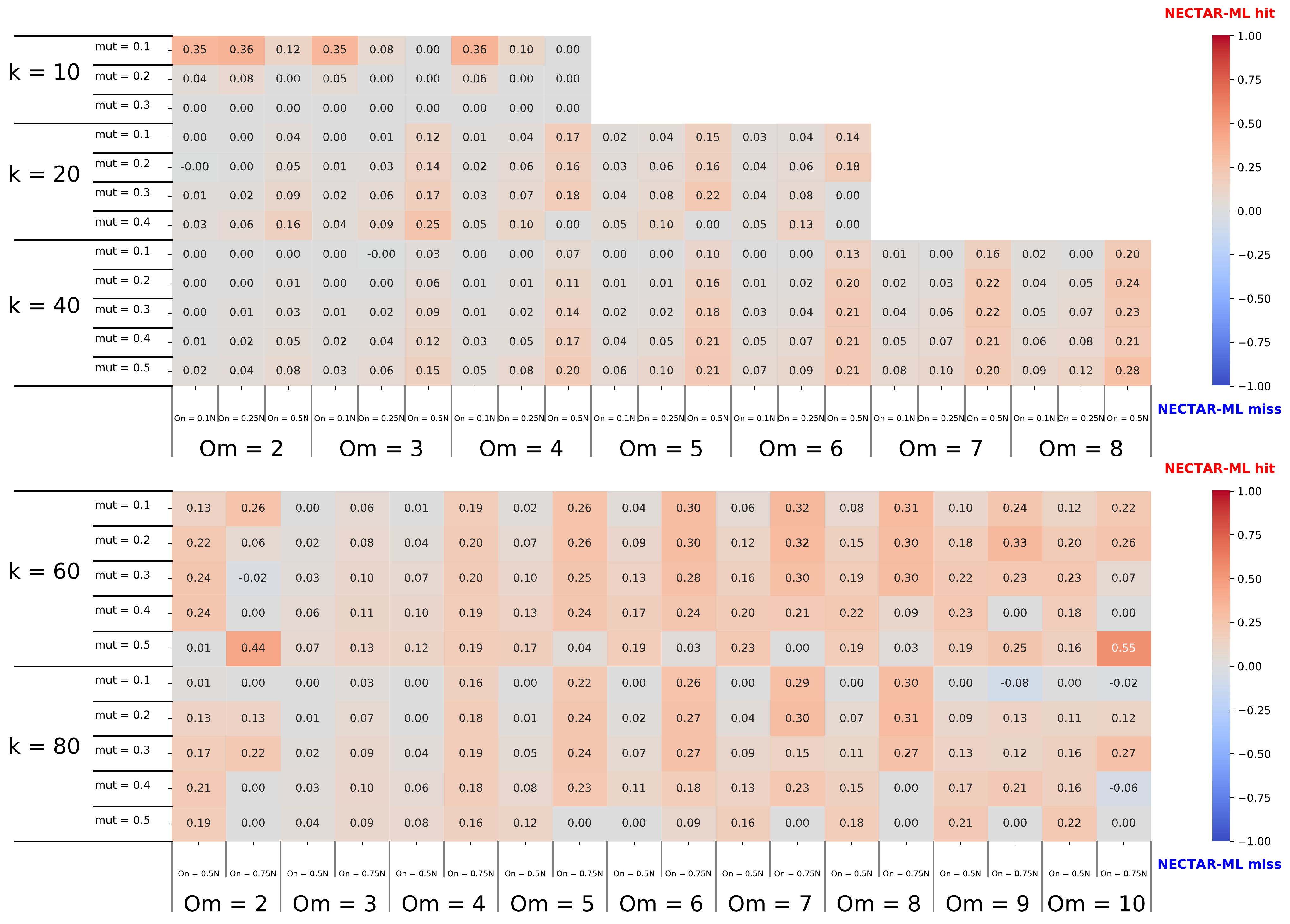}
\caption{NECTAR vs NECTAR-ML competitive analysis on synthetic networks using the Average F1 metrics model.} 
 \label{fig:compAnalysis_nectar_nectar_ml_AverageF1} 
\end{figure*}

\noindent \textbf{Synthetic Networks:} Fig. \ref{fig:compAnalysis_nectar_nectar_prime} presents the competitive analysis of Cohen et al's NECTAR\cite {NECTAR2017Cohen} and NECTAR-ML algorithms using the average metrics for labelling. Competitive analysis figures using ONMI, Omega-index and average F1 metrics follow similar trends and are presented in the Appendix (Figs. \ref{fig:compAnalysis_nectar_nectar_ml_ONMI}-\ref {fig:compAnalysis_nectar_nectar_ml_AverageF1}). In the figure, each cell in the heatmap represents a certain LFR configuration which is comprised out of four variables, $k$ $O_n$, $O_m$ and $mut$. The networks which match the cell's configuration are given a calculated weight according to Equation \eqref{eq:weight}. The value of the cell is calculated as follows. First, a total sum variable is initiaized to 0. Then, for each network, If NECTAR-ML selects the objective function correctly while NECTAR does not, the calculated weight value of the network is added to the total sum. If NECTAR selects the objective function correctly while NECTAR-ML does not, the calculated weight value is subtracted from the sum. If both algorithms are wrong/correct, no value is added nor subtracted from/to the sum. Since there are multiple networks per configuration, in order to normalise the value between -1 to 1, the summed value is divided by the number of networks which comprised the sum. The cell's color and intensity, with respect to the cell's value, are defined by the colour palette which is placed at the right of the heatmap. Therefore, a red colour indicates that for a specific configuration, NECTAR-ML provided better selections than NECTAR while a blue colour indicates the opposite. A grey colour indicates that both algorithms provided similar results. It can be seen that, except for only a few minor cases, NECTAR-ML is on average, equal to or better than the NECTAR algorithm. NECTAR-ML's better selections of the objective functions are more noticeable for the configurations containing $O_n$ values equal to 50\% and 75\% of the number of total nodes. In addition, we can see that this pattern is being strengthened with the increase of $k$ and $O_m$ values, meaning that NECTAR-ML presents supremacy over NECTAR especially in dense networks, as it can be seen from the bottom table in the figure.

\begin{figure}[!t]
\includegraphics[width = 0.6 \linewidth]{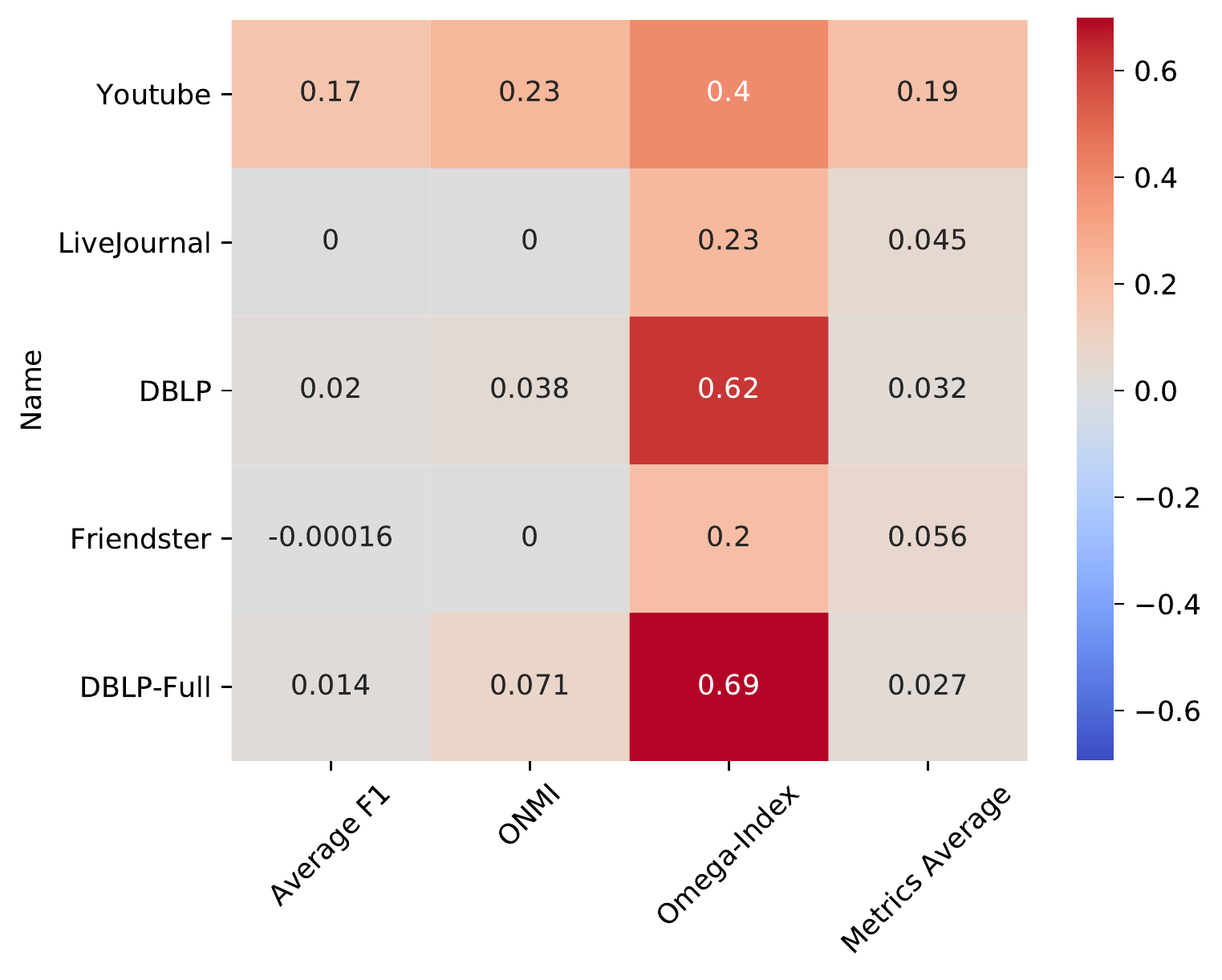}
\caption[NECTAR vs NECTAR-ML competitive analysis on real world networks per metric]{NECTAR vs NECTAR-ML competitive analysis on real world networks per metric.
}
\label{fig:rw_average_metrics} 
\end{figure}

\noindent \textbf{Real-world networks:} Fig. \ref{fig:rw_average_metrics} presents the competitive analysis of the NECTAR\cite {NECTAR2017Cohen} and NECTAR-ML algorithms, with respect to all metric types. As in figure \ref {fig:compAnalysis_nectar_nectar_prime}, the colour of the cell, represent a hit/miss of NECTAR-ML vs NECTAR while its opacity is controlled by the network's weight. The Amazon and Amazon-Full networks shared the same algorithms selections for all metric cases and are therefore omitted from the heatmap. It can be seen that NECTAR-ML's selections are equal or better than NECTAR, apart from one selection whose importance is very low, as the weighted value of this network is practically zero (Average F1 metric over the Friendster network). NECTAR-ML's superior objective function selections are reflected both on low and high weights values of the networks and therefore positions NECTAR-ML as significantly superior to NECTAR over real-world networks, for all four metrics.

\subsection{Competitive Analysis to MOEA algorithms}
\label{Competitive-Analysis-to-MOEA-algorithms}

In this section we analyse and describe the results of an extensive competitive analysis we have conducted between NECTAR-ML and the two MOEA algorithms, presented in Section \ref{seq:related}. First, all three algorithms are compared using the same benchmark of networks due to the network size limitation of the EMOFM-DK algorithm. Then, NECTAR-ML is compared to the MCMOEA algorithm over large scale networks.

\noindent \textbf{NECTAR-ML vs MOEAs:} For the purpose of evaluating correctly EA algorithms, a suitable benchmark was selected. The benchmark reflects a common ground of all algorithms and, for that reason, some considerations were taken into account. 
First, due to the runtime limitation of EMOFM-DK, the number of total nodes which was selected for the comparison was set to 1000. This network size is much smaller than the large scale networks we used in order to evaluate NECTAR-ML, but it enabled us to compare the algorithms properly. Second, the evaluation of the clusterings output by the algorithms was made using the commonly-used metrics, ONMI \cite {Lancichinetti_2009}, Omega Index \cite {Linda1988Omega} and Average F1 score \cite {Yang2012Affiliation}. Since ONMI and gNMI are versions of NMI and both are suitable for overlapping clustering evaluation, as stated in Section \ref{seq:related}, we decided to continue to use the ONMI metric, in order to use it consistently all throughout this work. Third, evolutionary multiobjective optimization based approaches can obtain multiple solutions of overlapping community structures in a single run, while NECTAR-ML is producing only a single solution per objective function and $\beta$ value (10 values of $\beta$ are used). To mitigate this gap, we decided to compare MOEAs best score per metric, selected from the entire set of runs and clustering outputs to NECTAR-ML's best score per metric, retrieved using the predicted objective function over the $\beta$ values. Finally, the number of runs for a MOEA algorithm to be executed  is set to 30, which is the number of runs used by Wen et al. and Tian et al. in their competitive analysis. Both MOEA algorithms run with their default parameters values.
We created a dataset of 785 synthetic networks for the competitive analysis. Specifically, the LFR parameters used by \cite{Xie2013Comparative} are presented in table \ref {tab:synthticNetworksForMOEAsComparsion}. The parameters, which were used in both MCMOEA \cite{Wen2017MaximalClique} and EMOFM-DK \cite{Tian2020Fuzzy} competitive analysis to other algorithms, are contained in this set of parameters. Due to the fact that NECTAR-ML's learning based model was designed for large scale networks and the analysis of significantly smaller ones present a different view on the classes and features distributions, the model was re-trained using smaller network sizes. The network sizes which were selected for the training phase were 100, 300, 500, 1400, 1600, 1800 and 2000. 

\begin{table}[b]
\centering
\caption{Parameters used for synthetic networks creation using LFR. Few networks failed to be generated by the LFR benchmark and therefore are missing from the dataset.}
\label{tab:synthticNetworksForMOEAsComparsion}
\resizebox{\columnwidth}{!}{%
\begin{tabular}{ |c|c|c|c|c|c|c|} 
\hline
$n$ &No. of Networks & $k$ & $maxK$ & $O_n$ & $O_m$ & $mut$\\
\hline
\multirow{2}{2em}{1000} & 374 & \{10,20\} & 50 & \{0.1,0.25,0.5\}& \{2,3,4,5,6,7,8\} & \{0.1,0.2,0.3,0.4,0.5\} \\ 
 & 411 & 40 & \{50,100\} & \{0.1,0.25,0.5\} & \{2,3,4,5,6,7,8\} & \{0.1,0.2,0.3,0.4,0.5\} \\
 \hline
 
 
\end{tabular}
}
\end{table}

All networks sizes share the same configurations as the 1000 network size configurations. The labels distribution, for a total of 5,918 synthetic networks, with and without consideration of the networks weight, is displayed in Table \ref{tab:SmallNetworksLabelsDistribution}.
 
\begin{table}[t]
\caption[Small scale synthetic networks labels distribution]{Objective functions labels distribution for small scale synthetic networks of size 100-2K with and without consideration of the networks weight. }
\label{tab:SmallNetworksLabelsDistribution}
\begin{center}
\begin{tabular}{| l | c | c | c | c |}
\hline
& \multicolumn{2}{c |}{Unweighted} & \multicolumn{2}{c |}{Weighted}\\
\cline{2-5}
Metric         &$Q^E$  &WOCC    &$Q^E$ &WOCC \\
\hline\hline

ONMI         	   &61\% 	&39\%   &56\%  &44\%   \\ \hline
Average F1         &61\% 	&39\%   &58\%  &42\%   \\ \hline
Omega-Index        &60\% 	&40\%   &50\%  &50\%   \\ \hline
Metrics Average    &60\% 	&40\%   &54\%   &46\%    \\ \hline

\end{tabular}
\end{center}
\end{table}

Table \ref{tab:SmallSyntheticResults} presents the results of the model evaluation on the test data for all metric types. The 1\textsuperscript{st} column presents the balanced accuracy results for the 5-folds cross-validation runs. The maximum standard deviation for the folds is less than 0.014. The remaining columns present the results for the networks test set. The 2\textsuperscript{nd} column present the balanced accuracy result and the last two columns present the recall results for each of the objective functions. All models are using the default threshold value of 0.5 to select between the objective functions. As can be seen, the balanced average achieved over the test set of the networks of size 1000 is close to 92\%. Also, the 5-folds cross validation results present similar results and have a low standard deviation of less than 0.014 for all metrics. This indicates that the models are not overfitted.

\begin{table}[!t]
\centering
\caption{Synthetic networks results for the classifiers on networks of size 1000.
}
\label{tab:SmallSyntheticResults}
\begin{tabular}{| l | c | c | c | c|}
\hline
Metric &Balanced Accuracy                    &Balanced Accuracy    &Recall - $Q^E$ & Recall - WOCC  \\ 
       &5-Folds       &Test Set     &Test Set          &Test Set\\
\hline\hline

ONMI          & 0.887  & 0.920 & 0.928 & 0.911 \\
Average F1    & 0.922  & 0.924 & 0.936 & 0.912 \\
Omega-Index   & 0.876  & 0.897 & 0.910  & 0.883 \\
Metrics Average & 0.917 & 0.933 & 0.935 & 0.931 \\
\hline

\textbf {Overall Average} & \textbf{0.901}  & \textbf{0.919} & \textbf{0.927} & \textbf{0.909} \\
\hline

 
\end{tabular}
\end{table}

\begin{figure}[H]
\includegraphics[width = 0.6 \linewidth]{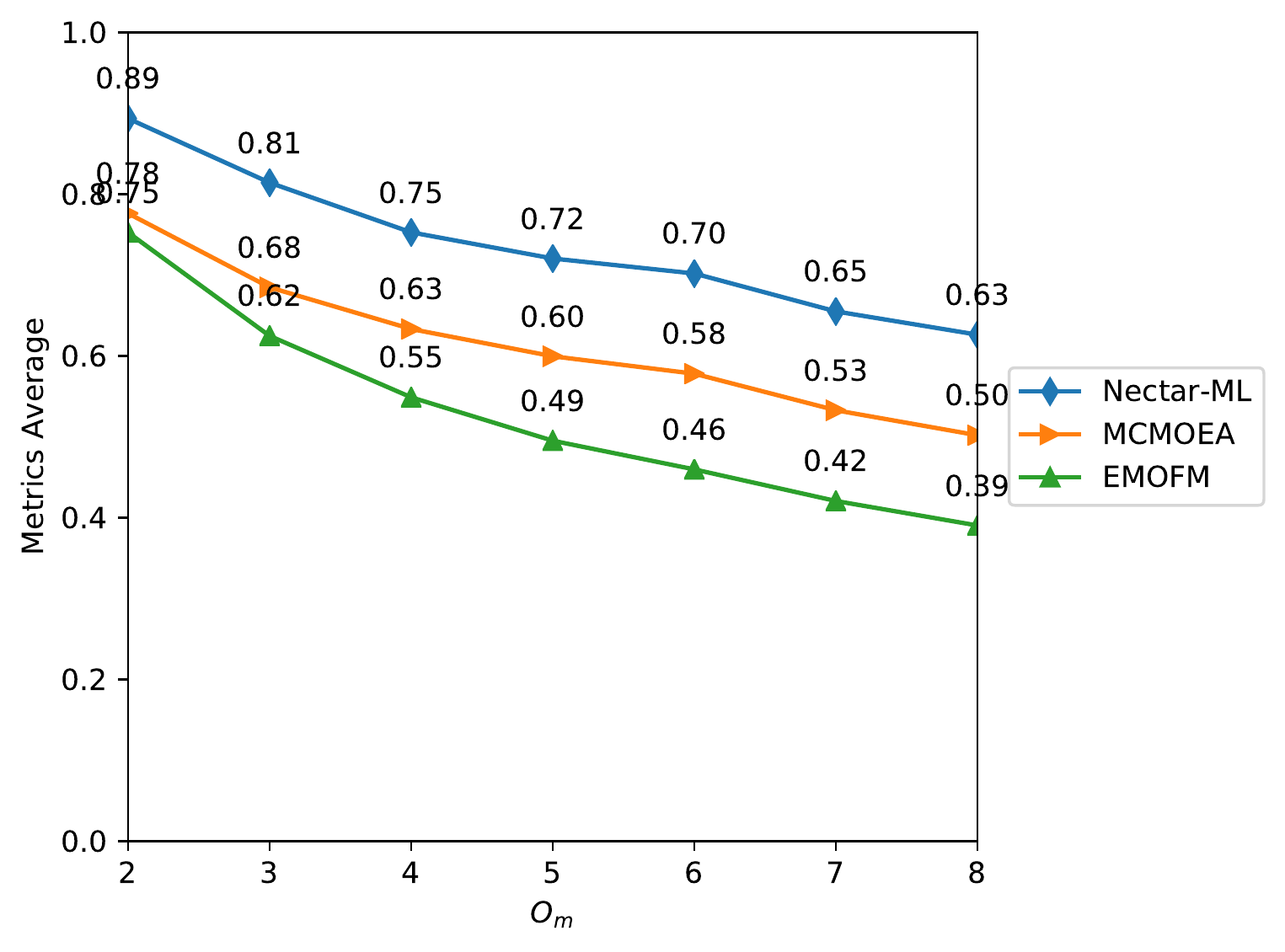}
\caption{The average performance of the average metrics value obtained by the three approaches on 785 networks, as a function of $O_m$.}
\label{fig:compAnalysis_moea_average} 
\end{figure}

Using our dedicated framework, we were able to process, for the EMOFM-DK and MCMOEA algorithms, approximately 9,891,000 and 1,177,500 clustering outputs respectively, as there were 785 compared networks, 30 runs per network and 420 clustering outputs for EMOFM-DK and 50 clustering outputs for MCMOEA per run. A timeframe of 48 hours was given for the MOEA algorithms to complete their runs per network. Fig. \ref{fig:compAnalysis_moea_average} presents the average performance of the algorithms in terms of average value of the three knowns metrics for all 785 networks. As can be seen in Fig. \ref{fig:compAnalysis_moea_average}, NECTAR-ML outperformed both algorithms,  while MCMOEA is the second-best, with a score lower than NECTAR-ML's by approximately 17\%. EMOFM-DK comes last with a score lower than NECTAR-ML's by approximately 30\%. Competitive analysis figures related to the ONMI, Omega-index and average F1 metrics follow the same trend and are presented in the Appendix. (Figs. \ref{fig:compAnalysis_moea_ONMI}-\ref{fig:compAnalysis_moea_AverageF1}).

\begin{figure} [t]
\includegraphics[width = 0.6 \linewidth]{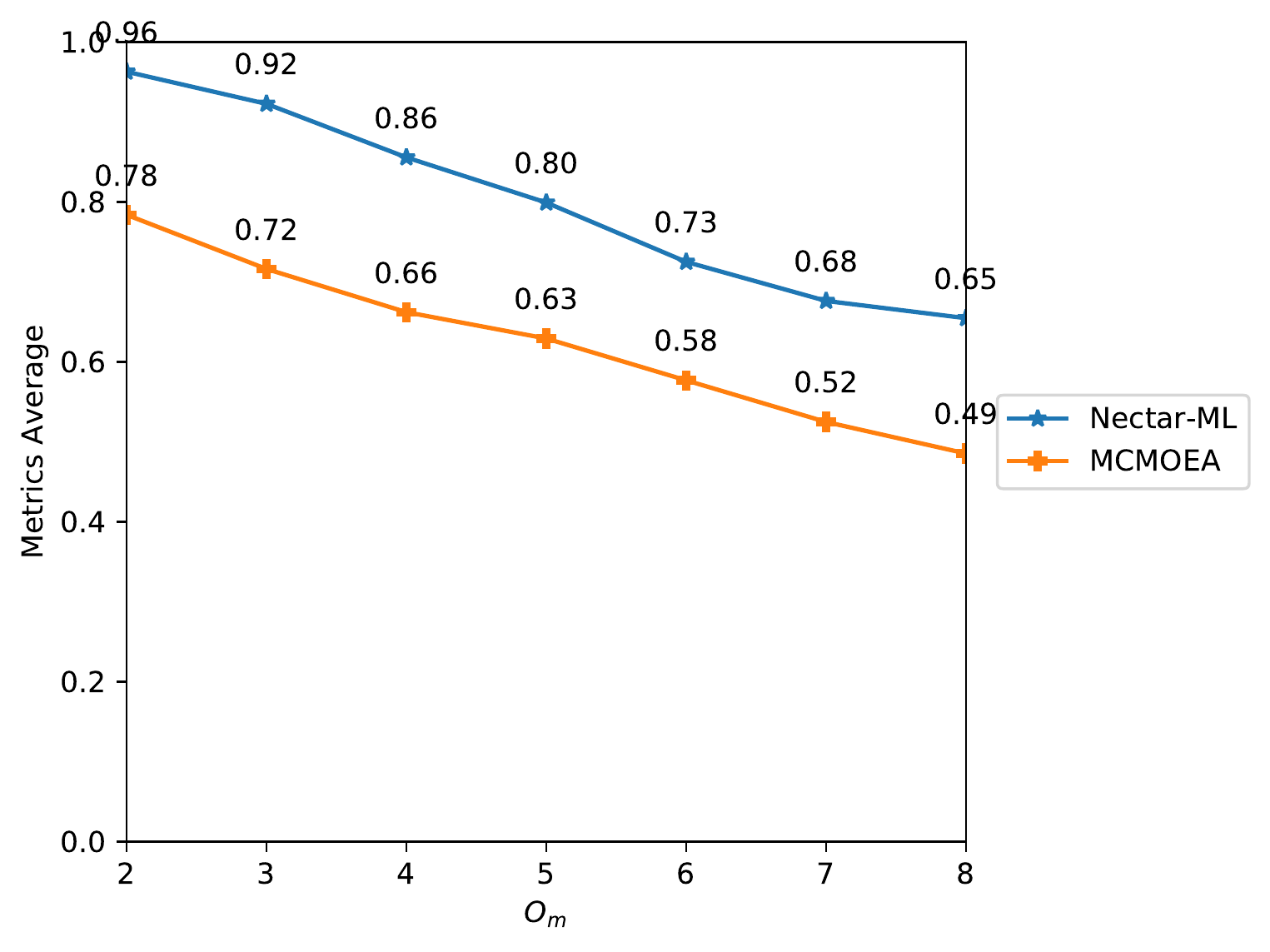}
\caption[NECTAR-ML and MCMOEA algorithms average metrics comparison on 10K networks]{The average performance of the average metrics value obtained by NECTAR-ML and MCMOEA algorithms on 195 large-scale networks, as a function of $O_m$. }
\label{fig:compAnalysis_mcmoea_average_10k} 
\end{figure}

\noindent \textbf{NECTAR-ML vs. MCMOEA :} We created a dataset of 195 synthetic networks for the large-scale networks competitive analysis. The selected LFR parameters extend the networks configuration used in MCMOEA \cite{Wen2017MaximalClique}, for the performance analysis of large-scale networks, as follows: The total number of nodes, $n$, is 10K, the average node degree, $k$, is set to 20. The number of overlapping nodes, $O_n$, is set to 10\%, 25\%, 50\%. The number of communities an overlapping node belongs to, $O_m$, is set to $\{2,\dots ,8\}$. The mixing parameter for the topology, $mut$, is set to values from the range $\{0.1,\dots ,0.5\}$. The maximum node degree $maxK$ is set to 50. Using the same platform used for the 1K competitive analysis, we were able to produce MCMOEA clustering outputs for the dataset. As for NECTAR-ML, since this dataset contains network configurations which are contained in the training set, the model was re-trained with these networks excluded. Then, the model was applied on the dataset. Fig. \ref{fig:compAnalysis_mcmoea_average_10k} presents the average performance of the algorithms in terms of average value of the three knowns metrics for all 195 networks. 

\begin{figure}[H]
\includegraphics[width = 0.6\linewidth]{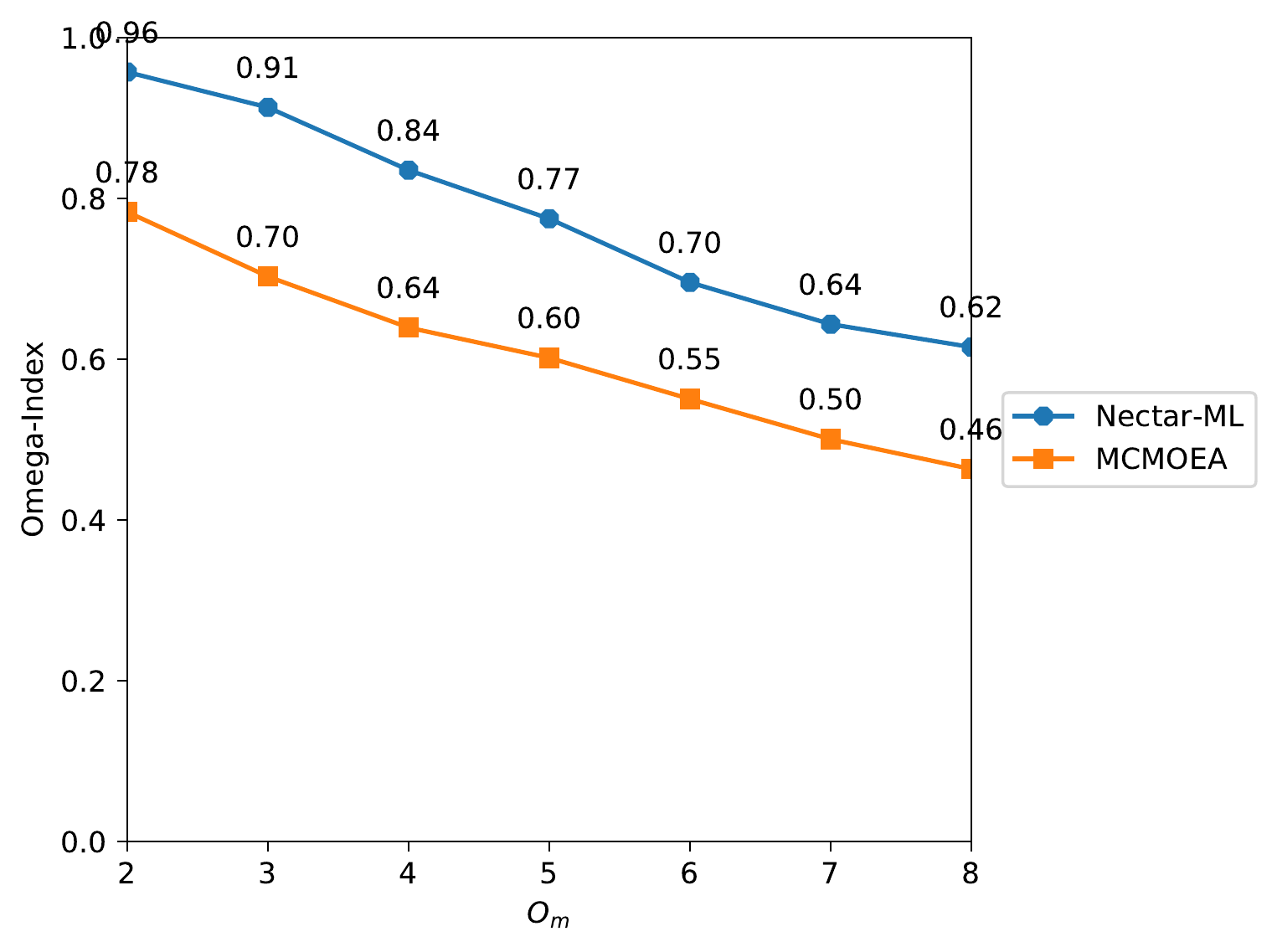}
\caption{The average performance of the Omega-Index metrics value obtained by NECTAR-ML and MCMOEA algorithms on 195 large-scale networks, as a function of $O_m$. }
\label{fig:compAnalysis_mcmoea_OmegaIndex_10k} 
\end{figure}

\begin{figure}[H]
\includegraphics[width = 0.6\linewidth]{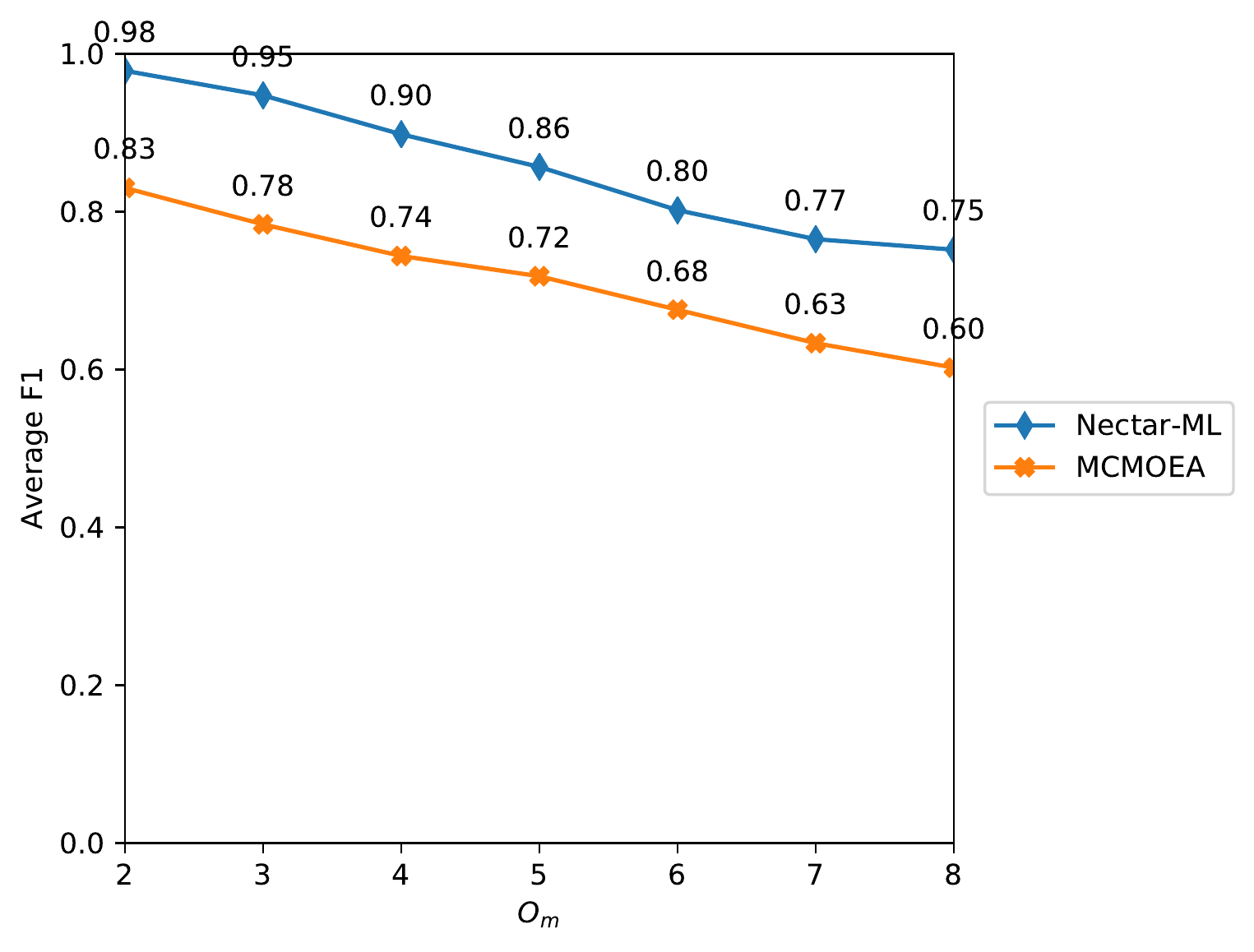}
\caption{The average performance of the Average F1 metrics value obtained by NECTAR-ML and MCMOEA algorithms on 195 large-scale networks, as a function of $O_m$. }
\label{fig:compAnalysis_mcmoea_AverageF1_10k} 
\end{figure}

As can be seen in Fig. \ref{fig:compAnalysis_mcmoea_average_10k}, NECTAR-ML outperformed MCMOEA, which obtained an averaged score lower than NECTAR-ML's by approximately 22\%. Competitive analysis figures related to The ONMI, Omega-index and average F1 metrics follow the same trend and are presented in the Appendix (Figs. \ref{fig:compAnalysis_mcmoea_ONMI_10k}-\ref{fig:compAnalysis_mcmoea_AverageF1_10k})

\section{Conclusion}
\label{sec:conc}

We introduced NECTAR-ML, an extension of the NECTAR algorithm that uses a machine-learning based model for automating the selection of the objective function by leveraging features based on the input graph. This model selects an objective function which maximizes the quality of the clusters computed by NECTAR. As several possible commonly-used metrics can be used to quantify this quality, a separate model was generated per each metric.

To train the model, we created a dataset of 15,755 synthetic networks, with various sizes and properties. We analyzed 3,933 synthetic and 7 real networks, measuring the quality of our models, aiming to dynamically select, based on the properties of the graph at hand, which objective function should be used. Our analysis shows that in approximately 90\% of the cases our model was able to successfully select the correct objective function to maximize the desired metric.
 
We conducted a competitive analysis of NECTAR and NECTAR-ML. NECTAR-ML was proven to be superior over NECTAR as it significantly outperformed NECTAR's ability to select the best objective function out of $Q^E$ and WOCC. In addition, we have conducted an extensive competitive analysis of NECTAR-ML and two additional state-of-the-art multi-objective algorithms based on the MOEA approach for the multiobjective optimization problem (MOP). NECTAR-ML outperformed both algorithms in terms of average detection quality. Multiobjective EAs (MOEAs) are considered to be the most popular approach to solve MOP and the fact that NECTAR-ML significantly outperforms them demonstrates the effectiveness of ML-based objective function selection.

This work can be extended in several ways. First, by using multi-class rather than binary classification, NECTAR-ML can be trained to select from a broader set of objective functions, possibly based on additional structural graph properties. As proposed in \cite {NECTAR2017Cohen} for the NECTAR algorithm, search heuristics that target some weighted average of several objective functions, rather than selecting just one of them, might further improve performance.

 Additionally, despite the fact that most community detection algorithms require at least one user-provided parameter, decreasing their number makes it easier to use the algorithm. In future work, we will attempt to do so by learning them from the input graph's structural features as well. For instance, predicting the optimal value for the $\beta$  parameter could be addressed as a regression problem.

\bibliographystyle{splncs04}
\bibliography{refs}
\phantomsection
\vskip -2\baselineskip plus -1fil 







\section{Appendix}

\subsection{Supplementary Tables}

\begin{table}[H]
\centering
\scriptsize
\caption{Parameters used for synthetic networks creation using LFR. Few networks failed to be generated by the LFR benchmark and therefore are missing from the dataset.}
\label{tab:synthticNetworks}
\begin{tabular}{ |c|c|c|c|c|c|c|} 
\hline
$n$ & No. of Networks & $k$ & $maxK$ & $O_n$ & $O_m$ & $mut$\\
  \hline
  \multirow{4}{3em}{10000} & 51 & 10 & 50 & \{0.1,0.25,0.5\} & \{2,3,4\} & \{0.1,0.2,0.3\}\\ 
  & 120 & 20 & 50 & \{0.1,0.25,0.5\} & \{2,3,4,5,6\} & \{0.1,0.2,0.3,0.4\} \\ 
  & 424 & 40 & \{50,100\} & \{0.1,0.25,0.5\} & \{2,3,4,5,6,7,8\} & \{0.1,0.2,0.3,0.4,0.5\} \\
  & 720 & \{60,80\} & \{100,120\} & \{0.5, 0.75\} & \{2,3,4,5,6,7,8,9,10\} & \{0.1,0.2,0.3,0.4,0.5\} \\ 
  \hline
  
  \hline
  \multirow{4}{3em}{15000} & 53 & 10 & 50 & \{0.1,0.25,0.5\} & \{2,3,4\} & \{0.1,0.2,0.3\}\\ 
  & 120 & 20 & 50 & \{0.1,0.25,0.5\} & \{2,3,4,5,6\} & \{0.1,0.2,0.3,0.4\} \\ 
  & 422 & 40 & \{50,100\} & \{0.1,0.25,0.5\} & \{2,3,4,5,6,7,8\} & \{0.1,0.2,0.3,0.4,0.5\} \\
  & 720 & \{60,80\} & \{100,120\} & \{0.5, 0.75\} & \{2,3,4,5,6,7,8,9,10\} & \{0.1,0.2,0.3,0.4,0.5\} \\
  \hline
  
  \hline
  \multirow{4}{3em}{15000} & 52 & 10 & 50 & \{0.1,0.25,0.5\} & \{2,3,4\} & \{0.1,0.2,0.3\}\\ 
  & 120 & 20 & 50 & \{0.1,0.25,0.5\} & \{2,3,4,5,6\} & \{0.1,0.2,0.3,0.4\} \\ 
  & 423 & 40 & \{50,100\} & \{0.1,0.25,0.5\} & \{2,3,4,5,6,7,8\} & \{0.1,0.2,0.3,0.4,0.5\} \\
  & 720 & \{60,80\} & \{100,120\} & \{0.5, 0.75\} & \{2,3,4,5,6,7,8,9,10\} & \{0.1,0.2,0.3,0.4,0.5\} \\
  \hline
  
  \hline
  \multirow{4}{3em}{20000} & 52 & 10 & 50 & \{0.1,0.25,0.5\} & \{2,3,4\} & \{0.1,0.2,0.3\}\\ 
  & 120 & 20 & 50 & \{0.1,0.25,0.5\} & \{2,3,4,5,6\} & \{0.1,0.2,0.3,0.4\} \\ 
  & 423 & 40 & \{50,100\} & \{0.1,0.25,0.5\} & \{2,3,4,5,6,7,8\} & \{0.1,0.2,0.3,0.4,0.5\} \\
  & 720 & \{60,80\} & \{100,120\} & \{0.5, 0.75\} & \{2,3,4,5,6,7,8,9,10\} & \{0.1,0.2,0.3,0.4,0.5\} \\
  \hline
  
  \hline
  \multirow{4}{3em}{25000} & 51 & 10 & 50 & \{0.1,0.25,0.5\} & \{2,3,4\} & \{0.1,0.2,0.3\}\\ 
  & 120 & 20 & 50 & \{0.1,0.25,0.5\} & \{2,3,4,5,6\} & \{0.1,0.2,0.3,0.4\} \\ 
  & 424 & 40 & \{50,100\} & \{0.1,0.25,0.5\} & \{2,3,4,5,6,7,8\} & \{0.1,0.2,0.3,0.4,0.5\} \\
  & 720 & \{60,80\} & \{100,120\} & \{0.5, 0.75\} & \{2,3,4,5,6,7,8,9,10\} & \{0.1,0.2,0.3,0.4,0.5\} \\
  \hline
  
  \hline
  \multirow{4}{3em}{30000} & 50 & 10 & 50 & \{0.1,0.25,0.5\} & \{2,3,4\} & \{0.1,0.2,0.3\}\\ 
  & 120 & 20 & 50 & \{0.1,0.25,0.5\} & \{2,3,4,5,6\} & \{0.1,0.2,0.3,0.4\} \\ 
  & 424 & 40 & \{50,100\} & \{0.1,0.25,0.5\} & \{2,3,4,5,6,7,8\} & \{0.1,0.2,0.3,0.4,0.5\} \\
  & 720 & \{60,80\} & \{100,120\} & \{0.5, 0.75\} & \{2,3,4,5,6,7,8,9,10\} & \{0.1,0.2,0.3,0.4,0.5\} \\
  \hline
  
  \hline
  \multirow{4}{3em}{35000} & 52 & 10 & 50 & \{0.1,0.25,0.5\} & \{2,3,4\} & \{0.1,0.2,0.3\}\\ 
  & 120 & 20 & 50 & \{0.1,0.25,0.5\} & \{2,3,4,5,6\} & \{0.1,0.2,0.3,0.4\} \\ 
  & 423 & 40 & \{50,100\} & \{0.1,0.25,0.5\} & \{2,3,4,5,6,7,8\} & \{0.1,0.2,0.3,0.4,0.5\} \\
  & 720 & \{60,80\} & \{100,120\} & \{0.5, 0.75\} & \{2,3,4,5,6,7,8,9,10\} & \{0.1,0.2,0.3,0.4,0.5\} \\
  \hline
  
  \hline
  \multirow{4}{3em}{40000} & 49 & 10 & 50 & \{0.1,0.25,0.5\} & \{2,3,4\} & \{0.1,0.2,0.3\}\\ 
  & 120 & 20 & 50 & \{0.1,0.25,0.5\} & \{2,3,4,5,6\} & \{0.1,0.2,0.3,0.4\} \\ 
  & 423 & 40 & \{50,100\} & \{0.1,0.25,0.5\} & \{2,3,4,5,6,7,8\} & \{0.1,0.2,0.3,0.4,0.5\} \\
  & 720 & \{60,80\} & \{100,120\} & \{0.5, 0.75\} & \{2,3,4,5,6,7,8,9,10\} & \{0.1,0.2,0.3,0.4,0.5\} \\
  \hline
  
  \hline
  \multirow{4}{3em}{45000} & 49 & 10 & 50 & \{0.1,0.25,0.5\} & \{2,3,4\} & \{0.1,0.2,0.3\}\\ 
  & 120 & 20 & 50 & \{0.1,0.25,0.5\} & \{2,3,4,5,6\} & \{0.1,0.2,0.3,0.4\} \\ 
  & 421 & 40 & \{50,100\} & \{0.1,0.25,0.5\} & \{2,3,4,5,6,7,8\} & \{0.1,0.2,0.3,0.4,0.5\} \\
  & 720 & \{60,80\} & \{100,120\} & \{0.5, 0.75\} & \{2,3,4,5,6,7,8,9,10\} & \{0.1,0.2,0.3,0.4,0.5\} \\
  \hline
  
  \hline
  \multirow{4}{3em}{50000} & 48 & 10 & 50 & \{0.1,0.25,0.5\} & \{2,3,4\} & \{0.1,0.2,0.3\}\\ 
  & 120 & 20 & 50 & \{0.1,0.25,0.5\} & \{2,3,4,5,6\} & \{0.1,0.2,0.3,0.4\} \\ 
  & 423 & 40 & \{50,100\} & \{0.1,0.25,0.5\} & \{2,3,4,5,6,7,8\} & \{0.1,0.2,0.3,0.4,0.5\} \\
  & 720 & \{60,80\} & \{100,120\} & \{0.5, 0.75\} & \{2,3,4,5,6,7,8,9,10\} & \{0.1,0.2,0.3,0.4,0.5\} \\
  \hline
  
  \hline
  \multirow{4}{3em}{55000} & 49 & 10 & 50 & \{0.1,0.25,0.5\} & \{2,3,4\} & \{0.1,0.2,0.3\}\\ 
  & 120 & 20 & 50 & \{0.1,0.25,0.5\} & \{2,3,4,5,6\} & \{0.1,0.2,0.3,0.4\} \\ 
  & 422 & 40 & \{50,100\} & \{0.1,0.25,0.5\} & \{2,3,4,5,6,7,8\} & \{0.1,0.2,0.3,0.4,0.5\} \\
  & 720 & \{60,80\} & \{100,120\} & \{0.5, 0.75\} & \{2,3,4,5,6,7,8,9,10\} & \{0.1,0.2,0.3,0.4,0.5\} \\
  \hline
  
  \hline
  \multirow{4}{3em}{60000} & 50 & 10 & 50 & \{0.1,0.25,0.5\} & \{2,3,4\} & \{0.1,0.2,0.3\}\\ 
  & 120 & 20 & 50 & \{0.1,0.25,0.5\} & \{2,3,4,5,6\} & \{0.1,0.2,0.3,0.4\} \\ 
  & 424 & 40 & \{50,100\} & \{0.1,0.25,0.5\} & \{2,3,4,5,6,7,8\} & \{0.1,0.2,0.3,0.4,0.5\} \\
  & 720 & \{60,80\} & \{100,120\} & \{0.5, 0.75\} & \{2,3,4,5,6,7,8,9,10\} & \{0.1,0.2,0.3,0.4,0.5\} \\
  \hline
  
  \hline
  \multirow{4}{3em}{65000} & 47 & 10 & 50 & \{0.1,0.25,0.5\} & \{2,3,4\} & \{0.1,0.2,0.3\}\\ 
  & 120 & 20 & 50 & \{0.1,0.25,0.5\} & \{2,3,4,5,6\} & \{0.1,0.2,0.3,0.4\} \\ 
  & 421 & 40 & \{50,100\} & \{0.1,0.25,0.5\} & \{2,3,4,5,6,7,8\} & \{0.1,0.2,0.3,0.4,0.5\} \\
  & 720 & \{60,80\} & \{100,120\} & \{0.5, 0.75\} & \{2,3,4,5,6,7,8,9,10\} & \{0.1,0.2,0.3,0.4,0.5\} \\
  \hline

\end{tabular}
\end{table}


\subsection{Supplementary Figures}

\begin{figure}[H]
\includegraphics[width = \linewidth]{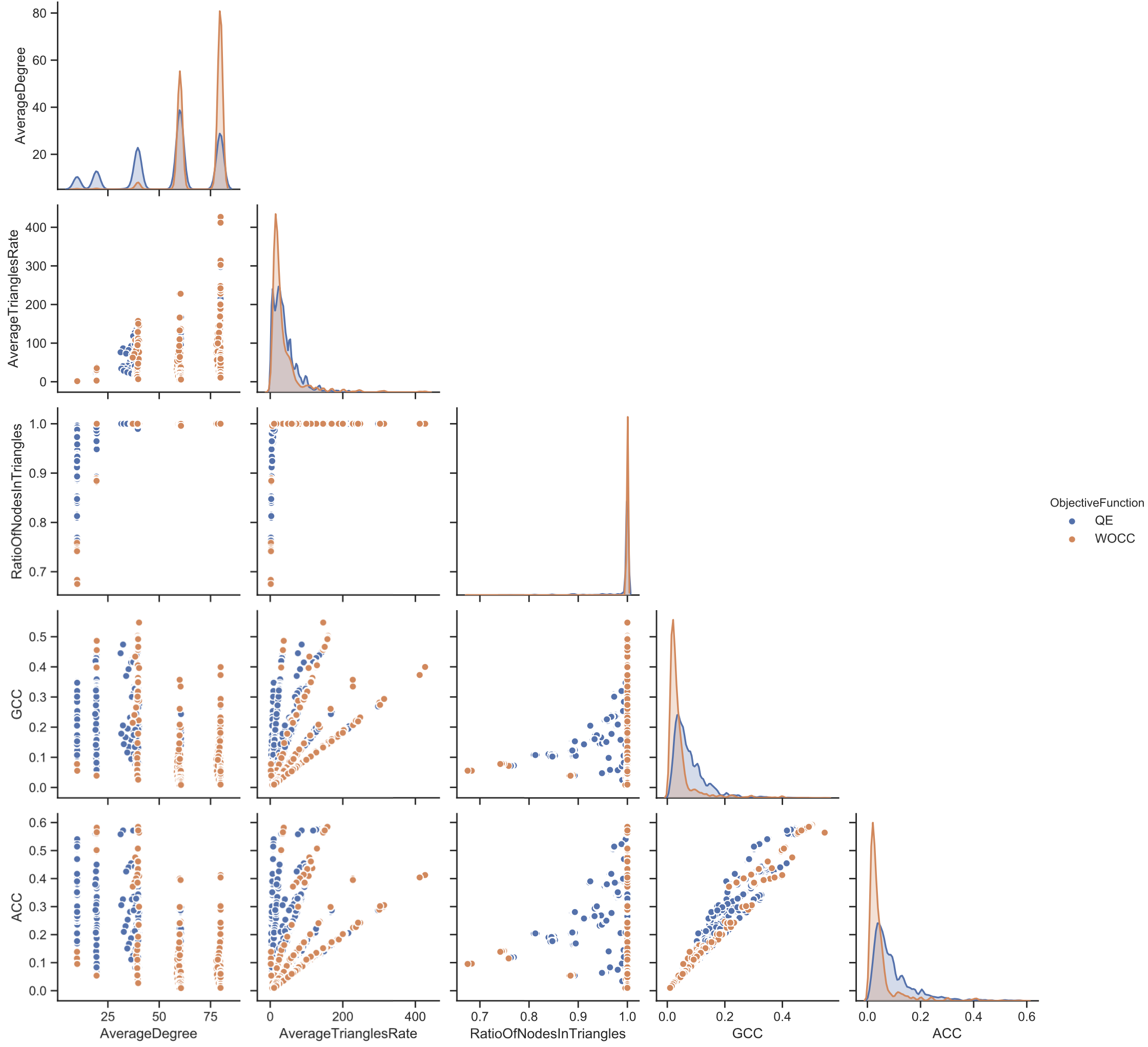}
\caption{Information Gain illustration for the ONMI metrics model.}
\label{fig:IG_ONMI} 
\end{figure}
\begin{figure}[H]
\includegraphics[width = \linewidth]{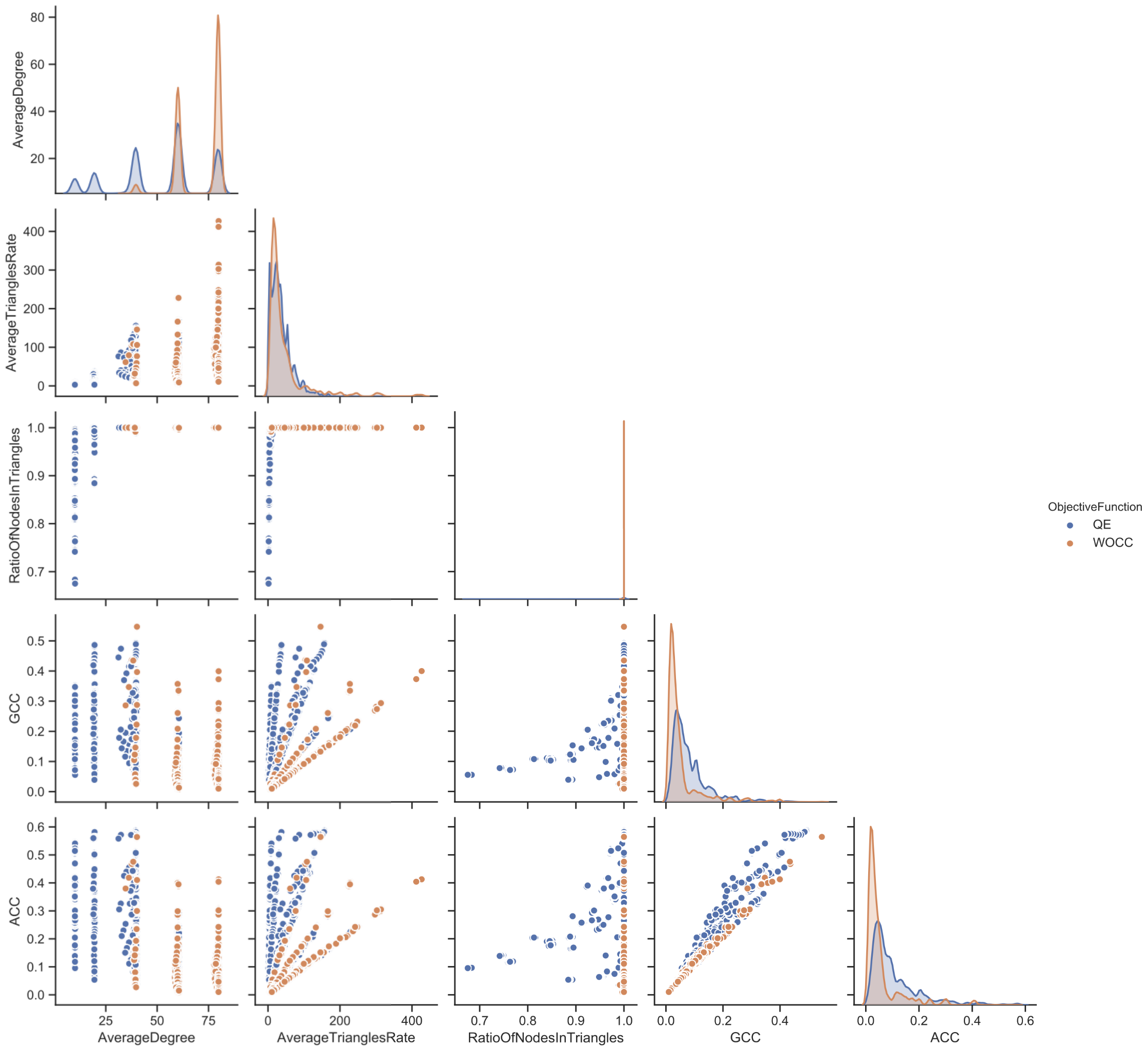}
\caption{Information Gain illustration for the Omega-Index metrics model.}
\label{fig:IG_OmegaIndex} 
\end{figure}
\begin{figure}[H]
\includegraphics[width = \linewidth]{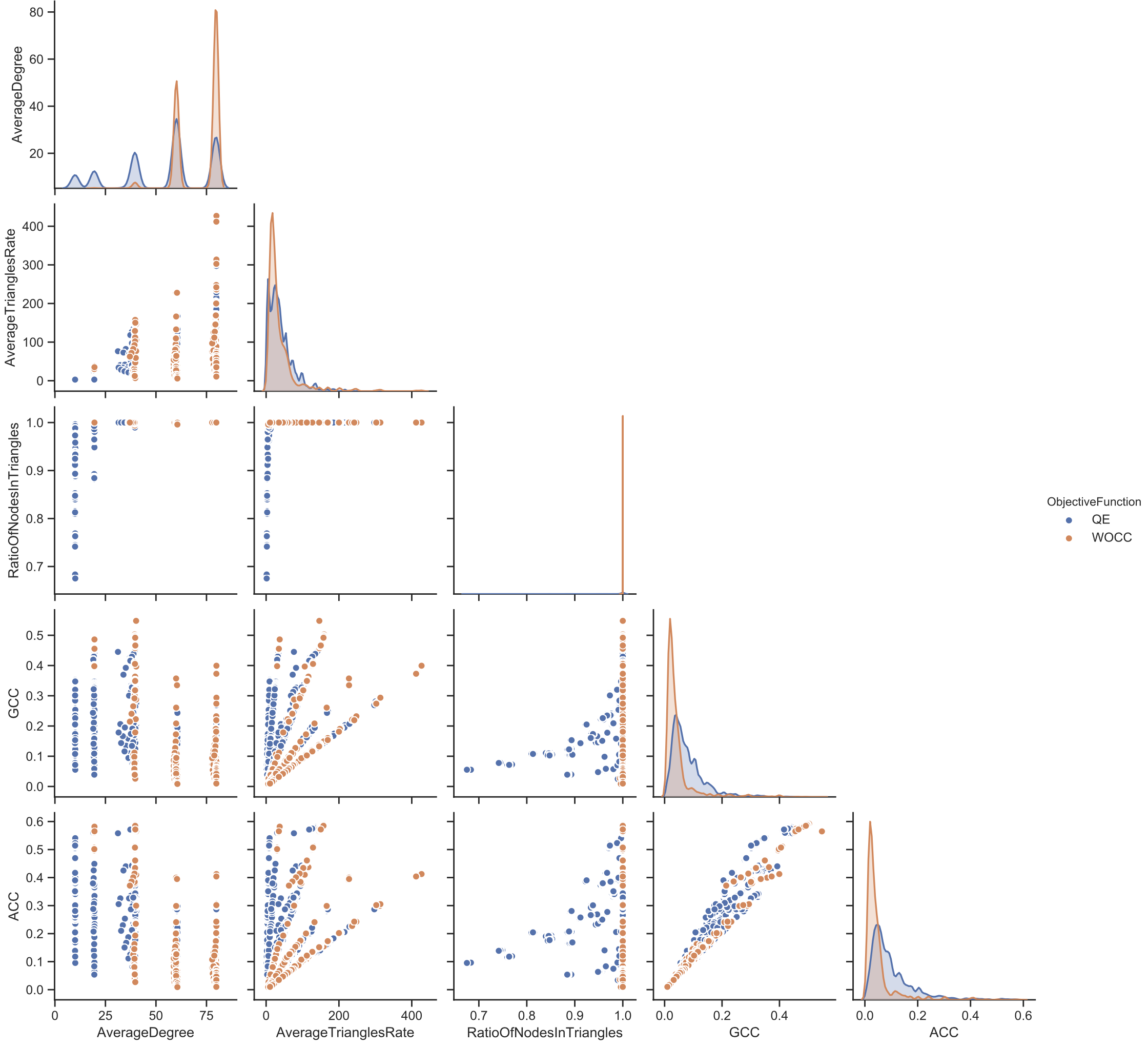}
\caption{Information Gain illustration for the Average F1 metrics model.}
\label{fig:IG_AverageF1} 
\end{figure}

\begin{figure*}
\centering
\includegraphics[width = \linewidth]{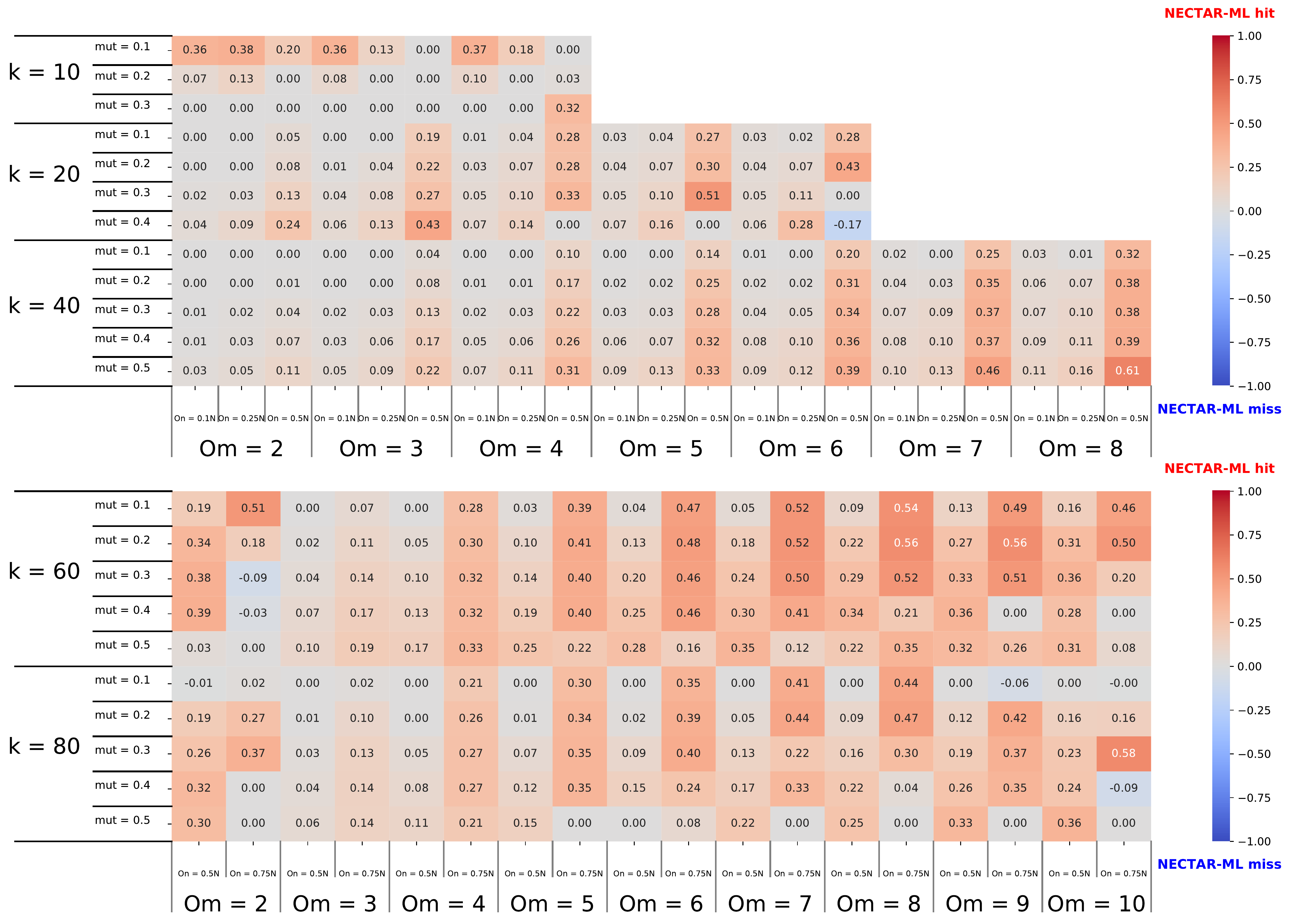}
\caption{NECTAR vs NECTAR-ML competitive analysis on synthetic networks using the ONMI metrics model.} 
 \label{fig:compAnalysis_nectar_nectar_ml_ONMI} 
\end{figure*}

\begin{figure}[H]
\includegraphics[width = 0.6\linewidth]{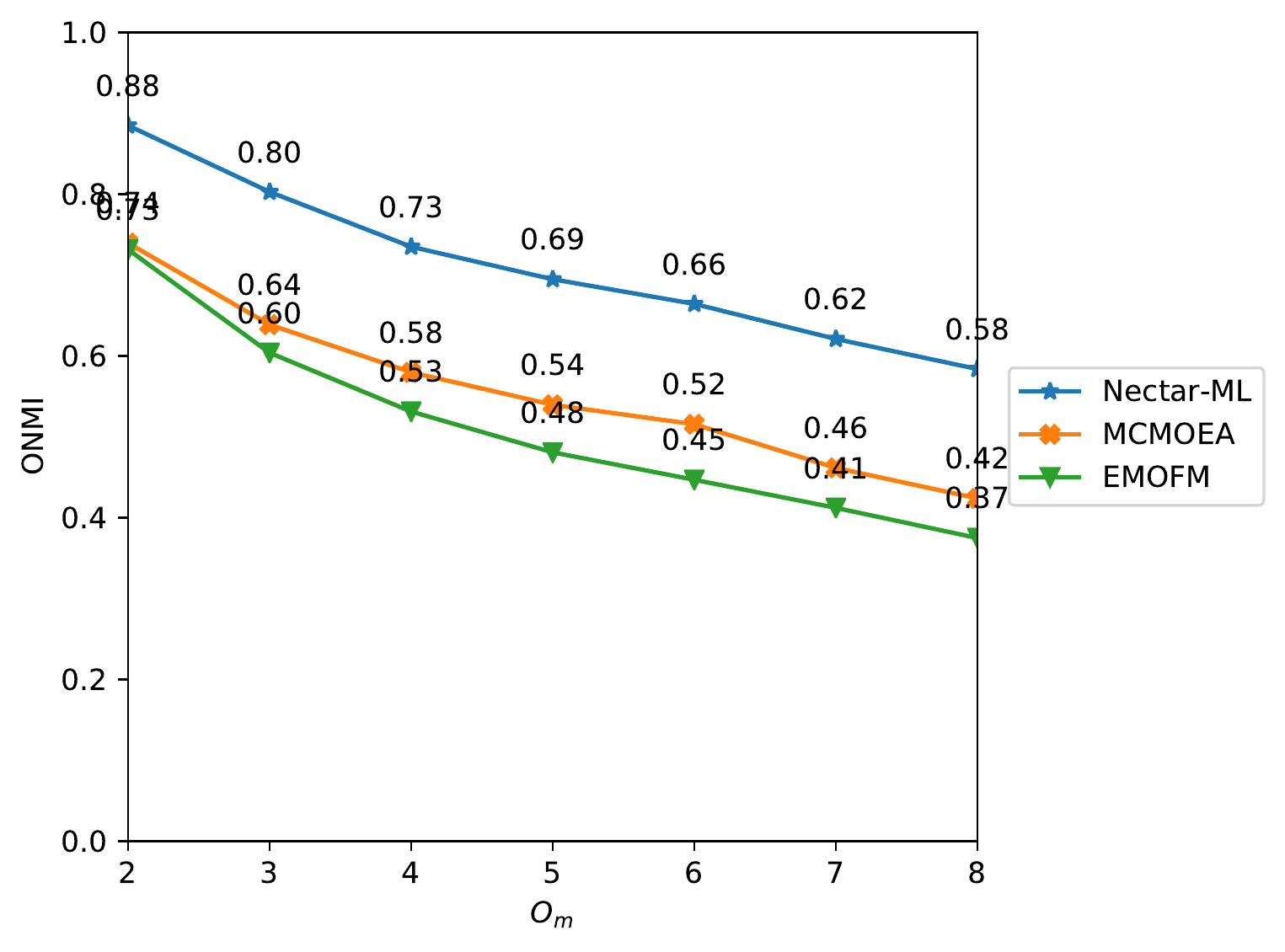}
\caption[NECTAR-ML and MOEA algorithms ONMI metrics comparison on 1K networks]{The average performance of the ONMI metrics value obtained by the three approaches on 785 networks, as a function of $O_m$.}
\label{fig:compAnalysis_moea_ONMI} 
\end{figure}
\begin{figure}[H]
\includegraphics[width = 0.6\linewidth]{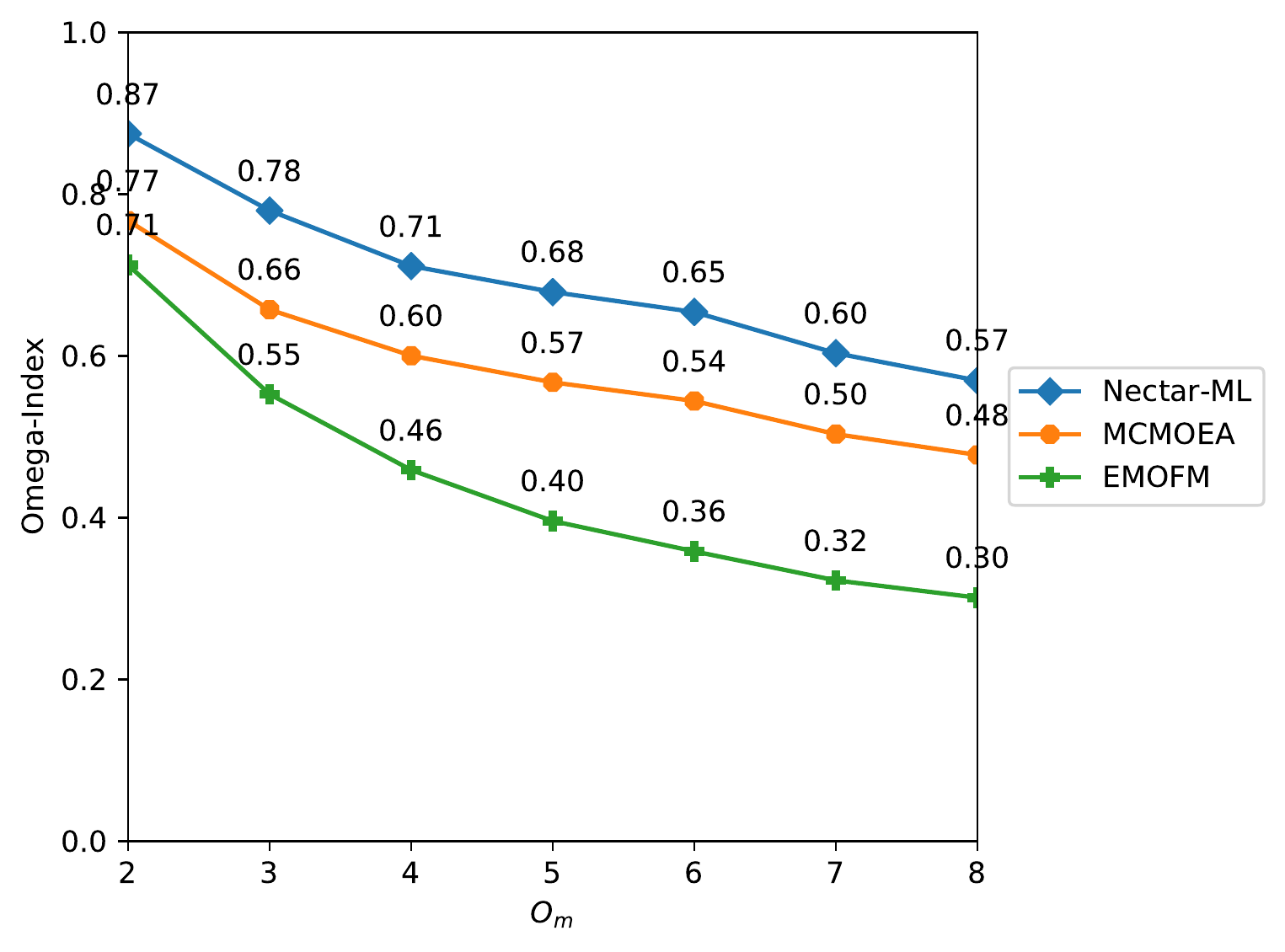}
\caption[NECTAR-ML and MOEA algorithms Omega-Index metrics comparison on 1K networks]{The average performance of the Omega-Index metrics value obtained by the three approaches on 785 networks, as a function of $O_m$.}
\label{fig:compAnalysis_moea_OmegaIndex} 
\end{figure}
\begin{figure}[H]
\includegraphics[width = 0.6\linewidth]{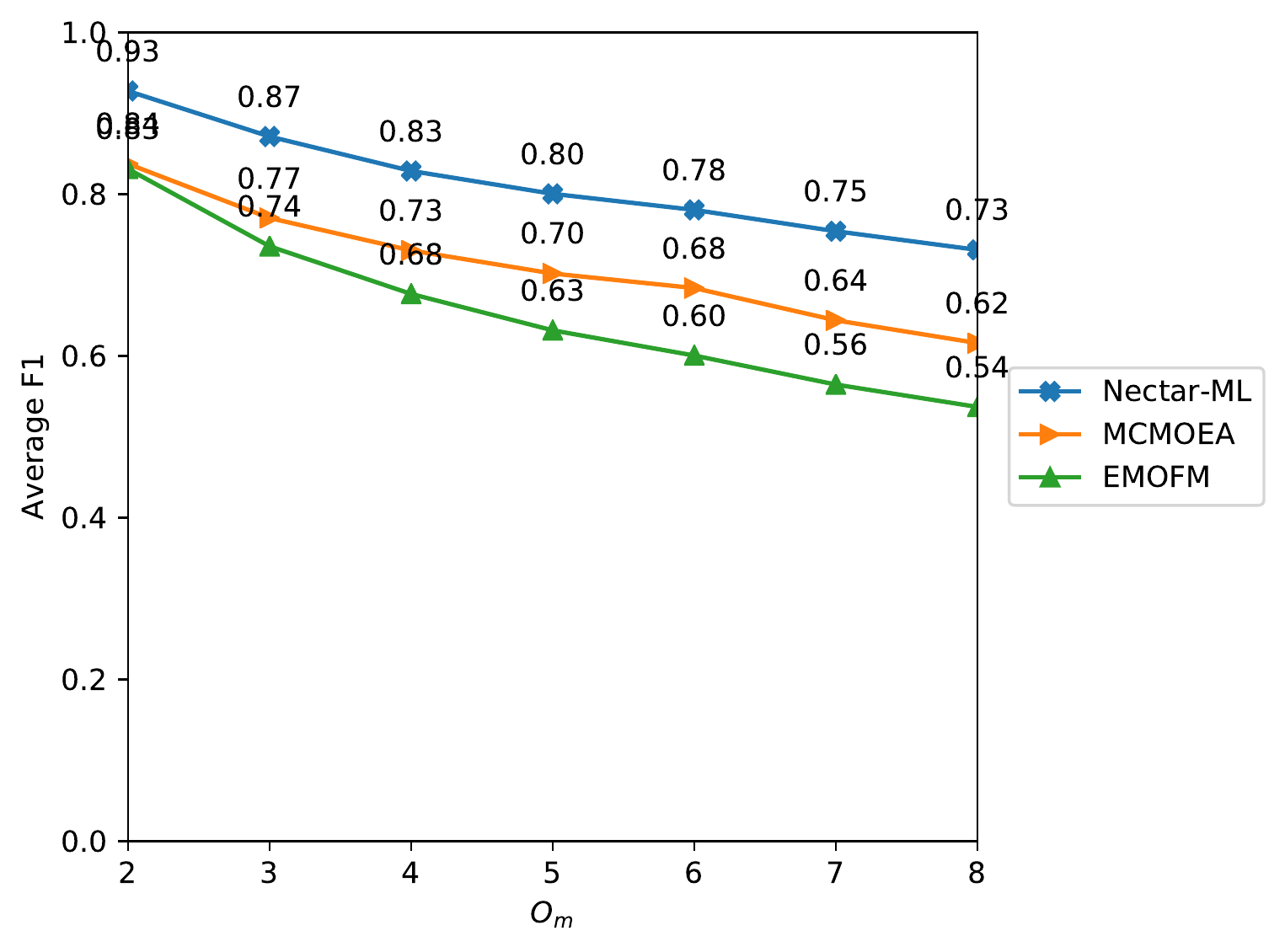}
\caption[NECTAR-ML and MOEA algorithms Average F1 metrics comparison on 1K networks]{The average performance of the Average F1 metrics value obtained by the three approaches on 785 networks, as a function of $O_m$.}
\label{fig:compAnalysis_moea_AverageF1} 
\end{figure}

\begin{figure}[H]
\includegraphics[width = 0.6\linewidth]{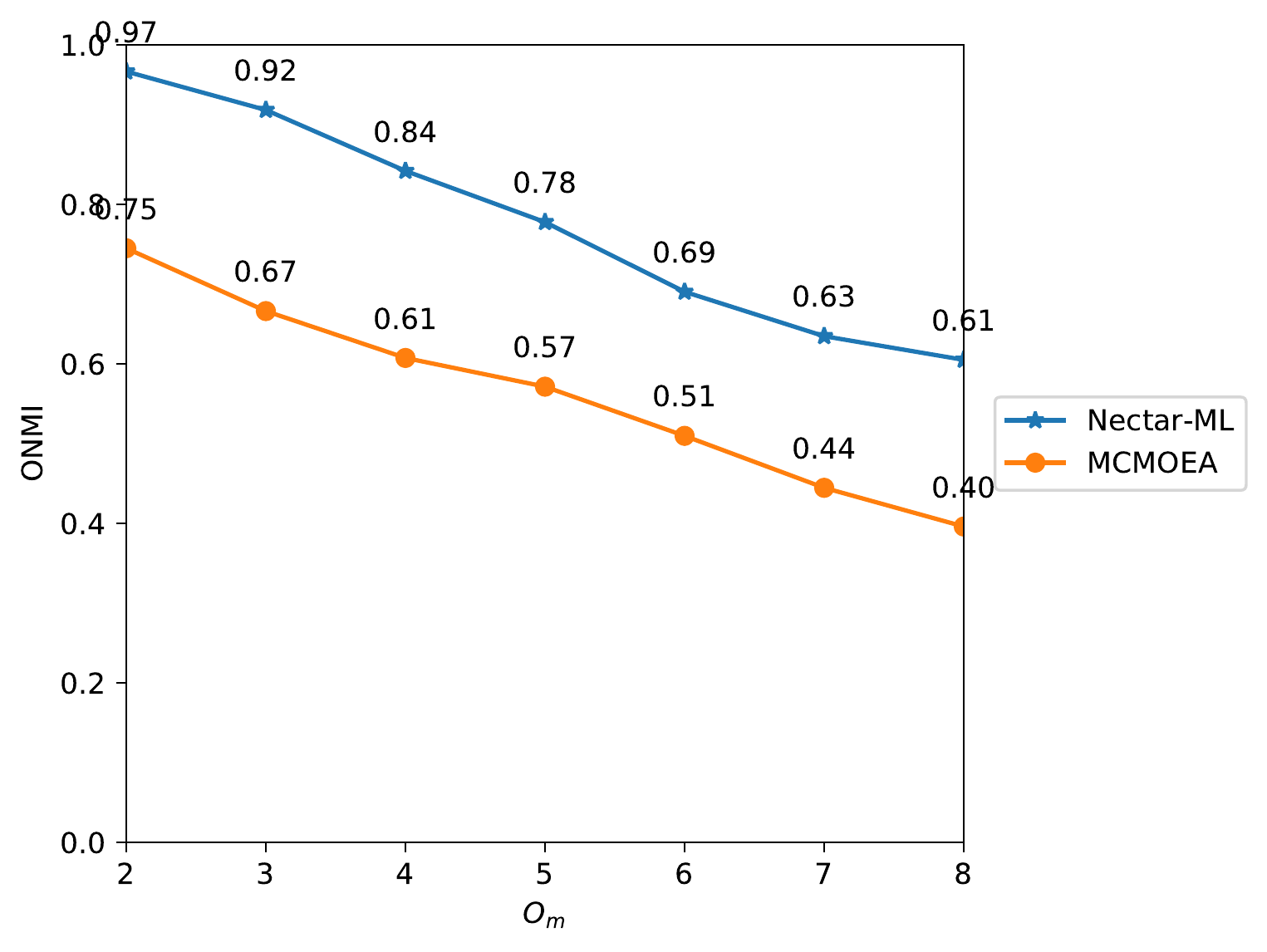}
\caption[NECTAR-ML and MCMOEA algorithms ONMI metrics comparison on 10K networks]{The average performance of the ONMI metrics value obtained by NECTAR-ML and MCMOEA algorithms on 195 large-scale networks, as a function of $O_m$. }
\label{fig:compAnalysis_mcmoea_ONMI_10k} 
\end{figure}

\end{document}